\documentclass[
twocolumn,
aps,
prb,
reprint,
superscriptaddress,
amsmath,amssymb
]{revtex4-1}
\usepackage[pdftex]{graphicx} % pdfLatex
\usepackage[hidelinks]{hyperref} % pdfLatex
\hypersetup{% hyperrefオプションリスト
 setpagesize=false,
 bookmarksnumbered=true,%
 bookmarksopen=true,%
 colorlinks=true,%
 linkcolor=blue,
 citecolor=blue,
}
\usepackage{braket}
\usepackage{times,multirow,amsfonts,bm,xspace,pifont}
\usepackage{soul}
\usepackage{ulem}
\usepackage{mathrsfs}
\begin{document}
\newcommand{\rr}{{\bm r}}
\newcommand{\q}{{\bm q}}
\renewcommand{\k}{{\bm k}}
\newcommand*\YY[1]{\textcolor{blue}{#1}}
\newcommand{\YYS}[1]{\textcolor{blue}{\sout{#1}}}
\newcommand*\YYWC[1]{\textcolor{green}{#1}}
\newcommand{\TK}[1]{{\color{red}{#1}}}
\newcommand*\TKS[1]{\textcolor{red}{\sout{#1}}}
\newcommand{\SK}[1]{{\color{magenta}{#1}}}
\newcommand{\MC}[1]{{\color{yellow}{#1}}}

\newcommand*\BdG{{\rm BdG}}

% Title of paper
\title{Quantum-geometry-induced anapole superconductivity% in parity-mixed superconductors
%General theory and quantum geometry for anapole superconductivity
}

%authors
\author{Taisei Kitamura}
\email[]{kitamura.taisei.67m@st.kyoto-u.ac.jp}
\affiliation{Department of Physics, Graduate School of Science, Kyoto University, Kyoto 606-8502, Japan}

\author{Shota Kanasugi}
%\email[]{}
%\affiliation{Department of Physics, Graduate School of Science, Kyoto University, Kyoto 606-8502, Japan}
\affiliation{Department of Physics, Graduate School of Science, Kyoto University, Kyoto 606-8502, Japan}

\author{Michiya Chazono}
%\email[]{}
\affiliation{Department of Physics, Graduate School of Science, Kyoto University, Kyoto 606-8502, Japan}

\author{Youichi Yanase}
%\email[]{yanase@scphys.kyoto-u.ac.jp}
\affiliation{Department of Physics, Graduate School of Science, Kyoto University, Kyoto 606-8502, Japan}

\date{\today}

\begin{abstract}
Anapole superconductivity recently proposed for multiband superconductors [\href{https://www.nature.com/articles/s42005-022-00804-7}{Commun. Phys. $\bm 5$, 39\ (2022)}] is a key feature of time-reversal ($\mathcal{T}$)-symmetry-broken polar superconductors.
The anapole moment was shown to arise from the asymmetric Bogoliubov spectrum, which induces a finite center of mass momenta of Cooper pairs at the zero magnetic field.
In this paper, we show an alternative mechanism of anapole superconductivity: the quantum geometry induces the anapole moment when the interband pairing and Berry connection are finite.
Thus, the anapole superconductivity is a ubiquitous feature of $\mathcal{T}$-broken multiband polar superconductors. 
Applying the theory to a minimal model of UTe$_2$, we demonstrate the quantum-geometry-induced anapole superconductivity.
Furthermore, we show the Bogoliubov Fermi surfaces (BFS) in an anapole superconducting state and predict an unusual temperature dependence of BFS due to the quantum geometry.
Experimental verification of these phenomena may clarify the superconducting state in UTe$_2$ and reveal the ubiquitous importance of quantum geometry in exotic superconductors.
\end{abstract}

\maketitle

\section{Introduction}

Parity-mixed superconductors, in which even- and odd-parity pairings coexist, are attracting much attention, as the parity-mixing phenomena are closely related to the space inversion ($\mathcal{P}$)-symmetry breaking. %in various fields~\cite{fu2015parity,kozii2015odd,sumita2020superconductivity,hiroi2018pyrochlore,schumann2020possible,bauer2012non,smidman2017superconductivity,fisher2022superconductivity,wu2010mixed,zhou2017fermion}.
Stimulated by the discovery of noncentrosymmetric superconductivity in heavy fermions and artificial heterostructures, time-reversal ($\mathcal{T}$)-symmetric parity-mixed pairing states such as the $s+p$-wave state have been investigated intensively~\cite{bauer2012non,smidman2017superconductivity}. 
%; the representative one is the $s+p$-wave pairing.
%
For a long time, studies focused on the crystals lacking the $\mathcal{P}$-symmetry allowing an antisymmetric spin-orbit coupling (ASOC). Consequently, the Rashba superconductor and the Ising superconductor have become fundamental concepts in condensed matter physics~\cite{bauer2012non,smidman2017superconductivity}.
%~\cite{fu2015parity,kozii2015odd,hiroi2018pyrochlore,sumita2020superconductivity,schumann2020possible}. 

On the other hand, centrosymmetric crystals were recently shown to be an intriguing platform of spontaneously $\mathcal{P}$-symmetry breaking superconductivity~\cite{sergienko2004mixed,wang2017topological,yang2020single}.
%, as followed:
%~\cite{fisher2022superconductivity}.%,wu2010mixed,zhou2017fermion}.
In the absence of the ASOC, additional $\mathcal{T}$-symmetry breaking is expected~\cite{wang2017topological,yang2020single,sergienko2004mixed} as %Spontaneous $\mathcal{T}$-symmetry breaking is realized by 
the ±$\pi$/2 phase difference between even- and odd-parity pairing potentials, such as the $s+ip$-wave pairing state, is energetically favored.
As a result, both of the $\mathcal{P}$- and $\mathcal{T}$-symmetry are broken while the combined $\mathcal{PT}$-symmetry is preserved.
The three-dimensional $s+ip$-wave pairing state in single-band superconductors was theoretically studied as a superconducting analog~\cite{goswami2014axioic,shiozaki2014dynamical,roy2020higher} of axion insulators~\cite{ryu2012electromagnetic,zi2013axion}. Such a paring state in Sr$_2$RuO$_4$ was theoretically proposed~\cite{scaffidi2020degeracy}.
Furthermore, recently discovered candidate for spin-triplet superconductor UTe$_2$~\cite{ran2019nearly,Aoki_review2022}
%which has a locally noncentrosymmetric crystal structure
is predicted to realize the $s+ip$-wave pairing state~\cite{ishizuka2021periodic}, as it is consistent with the experimentally observed multiple superconducting phases~\cite{braithwaite2019multiple,ran2020enhancement,lin2020tuning,knebel2020anisotorpy,aoki2020multiple,thomas2020evidence,thomas2021spatially} and multiple magnetic fluctuations~\cite{duan202incommensurate,valieska2021magnetic,butch2022symmetry,duan2021resonance,raymond2021feedback,Ambika2022}.

Clarification of the $\mathcal{PT}$-symmetric parity-mixed superconductivity has been awaited to uncover an exotic state of matter. 
However, properties of the $\mathcal{PT}$-symmetric parity-mixed superconductivity are almost unresolved. In particular, 
theoretical studies of \textit{multiband} superconductors have not been carried out except for Ref.~\onlinecite{kanasugi2022anapole}, although it is known that intriguing superconducting phenomena such as the intrinsic polar Kerr effect~\cite{wysokinski2012intrinsic,taylor2012intrinsic,gradhand2013kerr} and Bogoliubov Fermi surfaces (BFS)~\cite{agterberg2017bogoliubov,brydon2018bogoliubov} may appear from multiband properties.
In Ref.~\onlinecite{kanasugi2022anapole}, the anapole superconductivity was discussed as an exotic feature of the $\mathcal{PT}$-symmetric parity-mixed pairing state in multiband superconductors.
If some conditions are satisfied, an asymmetric Bogoliubov spectrum (BS) arises from the %$\mathcal{PT}$-symmetric parity-mixed 
interband pairing~\cite{kanasugi2022anapole}.
When the symmetry of superconductivity has a polar property, such as in the $A_{{\rm g}}+iB_{3{\rm u}}$ pairing state proposed for UTe$_2$~\cite{ishizuka2021periodic}, the asymmetric BS induces an effective anapole moment, which is defined as the first-order coefficient of the free energy in terms of the center of mass momenta of Cooper pairs.
%which is called anapole superconductivity from the perspective of symmetry.
The anapole moment characterizes the anapole superconductivity as it does the anapole order in magnetic materials~\cite{spaldin2008toroidal,jeong2017time,murayam2021bond,watanabe2021chiral,ahn2020low} and nucleus~\cite{flambaum1984nuclear}. 

The anapole moment is a polar and $\mathcal{T}$-odd  vector~\cite{spaldin2008toroidal}, which shares the symmetry as the velocity and momentum. 
%which is the so-called anapole state whose concept is studied in various fields.
%Surprisingly the effective anapole moment induces the finite Cooper pairs center of mass moment
Therefore, it is not surprising that the effective anapole moment induces a finite center of mass momenta of Cooper pairs $\bm q$ even in the absence of the magnetic field. 
The mechanism of finite-$\bm q$ pairing is different from 
%while the finite center of mass momenta are conventionally induced by the magnetic field in 
the Fulde-Ferrell-Larkin-Ovchinnikov (FFLO) superconductivity~\cite{flude1964superconductivity,larkin1964nonuniform} and helical superconductivity~\cite{bauer2012non,smidman2017superconductivity}, which require a finite magnetic field. 
In contrast to the FFLO and helical superconductivity, the anapole superconductivity can be studied with avoiding experimental difficulties due to vortices induced by an external magnetic field. For instance, the anapole domain switching~\cite{kanasugi2022anapole}, superconducting piezoelectric effect~\cite{Chazono_SCPE,piezocomment}, and Josephson effect~\cite{kaur2005helical,josephcomment} may uncover intrinsic properties of anapole superconductivity.
Therefore, the anapole superconductivity may be the key to elucidating the $\mathcal{PT}$-symmetric parity-mixed pairing state, and it may realize and clarify the finite-$\bm q$ pairing state which has been searched for a long time~\cite{bauer2012non,smidman2017superconductivity,matsuda_shimahara_FFLO}.

In this paper, we show that the anapole superconductivity is a ubiquitous feature more than revealed in the previous paper~\cite{kanasugi2022anapole}, %mixed-parity pairing state in multiband superconductors, owing to 
considering the \textit{quantum geometry} %which has recently been 
extensively studied in various fields~\cite{marzari1997maximally,resta2011the,gao2014field,gao2019nonreciprocal,lapa2019semiclassical,Daido_quadrupole,julku2021excitations,julku2021quantum,solnyshkov2021quantum,liao2021experimental,ahn2020low,watanabe2021chiral,rhim2020quantum,kitamura2021thermodynamic}.
%The anapole moment for superconductors is defined as the first-order coefficient of the free energy in terms of $\bm q$, which is directly related to the finite-$\bm q$ Cooper pairing~\cite{kanasugi2022anapole}. %in the $\mathcal{PT}$-symmetric parity-mixed superconductors, 
Recently, an essential role of the quantum geometry in the superfluid weight, namely, the second-order derivative of the free energy, has been revealed~\cite{peotta2015superfluidity,liang2017band,torma2022superconductivity,kitamura2021superconductivity}. %~\cite{julku2016geometric,he2021geometry,hu2019geometric,julku2020superfluid,xie2020topology-bonded,peri2021fragile,rossi2021quantum,torma2021superfluidity,herzogarbeitman2021superfluid,tian2021evidence,kitamura2021superconductivity,lau2022universal,huhtinen2022revising}. 
Thus, it is naturally expected that the quantum geometry may be essential for the anapole superconductivity.

First, we provide a thorough formulation of the anapole moment based on the Bardeen-Cooper-Schrieffer (BCS) mean-field theory. %Then, we discuss the origin of anapole superconductivity using the Ginzburg-Landau theory.
%While previous literature focuses on the special case to derive the effective anapole moment~\cite{kanasugi2022anapole}, we derive the effective anapole moment in a more general case.
The obtained formula contains two terms; one is the geometric term and the other is the group velocity term. Only a part of the group velocity term was derived in the previous literature~\cite{kanasugi2022anapole}.
Based on the general two-band model with Kramers degeneracy, the microscopic origin of the geometric term is revealed to be the 
%$\mathcal{P}$-odd (or even) 
interband pairing and the Berry connection, while the group velocity term is induced by the asymmetric BS.
Then, applying the theory to a model of UTe$_2$, we demonstrate the quantum-geometry-induced anapole superconductivity.
Moreover, we show unique features of anapole superconductivity.
When the system has a small gap minimum as expected for UTe$_2$~\cite{Aoki_review2022}, the anapole moment induces the BFS. The BFS may show a reappearing behavior as decreasing the temperature, causing anomalies in density of states (DOS) and thermodynamic quantities.

\section{General formula for anapole moment}
An order parameter of the anapole superconductivity is the anapole moment which is defined by the first-order coefficient of the free energy with respect to $\bm q$. %in the $\mathcal{PT}$-symmetric parity mixed superconductor.
In the previous study~\cite{kanasugi2022anapole}, the anapole moment is derived only 
when the ${\bm k}$-derivative of normal-state Hamiltonian is proportional to the identity matrix, namely, $\partial_\mu H_{\bm k}  \propto\bm 1$.
%in the case for $\partial_\mu H_{\bm k}  \propto\bm 1$ using the GL theory and where $\bm 1$ is the identity matrix, $H_{\bm k}$ is the matrix representation of the normal-state Hamiltonian. 
We adopt the notation $\partial_\mu=\partial_{k_{\mu}}$, and $H_{\bm k}$ is the matrix representation of the single-particle Hamiltonian. 
Below, we formulate the anapole moment in the general case based on the BCS mean-field theory.

The normal state is assumed to be $\mathcal{P}$- and $\mathcal{T}$-symmetric, and therefore, $H_{\bm k} = U_{\mathcal{T}} H_{-\bm k}^{T}U_{\mathcal{T}}^\dagger$ is satisfied, where $U_{\mathcal{T}}=i\sigma_y\otimes \bm 1$ is the unitary part of the $\mathcal{T}$ operator with the Pauli matrix for the spin space $\sigma_{\mu}$ ($\mu = 0,x,y,z$).
Thus, the Bogoliubov-de Gennes (BdG) Hamiltonian for a finite-$\bm q$ pairing state can be written as (see Appendix~\ref{sec:ana_derivation})
\begin{align}
	\hat{H}^{\BdG} &= \frac{1}{2}\sum_{\bm k}\hat{\Psi}^\dagger_{\bm k,\bm q}H^{\BdG}_{\bm k,\bm q}\hat{\Psi}_{\bm k,\bm q},\\
	H^{\BdG}_{\bm k,\bm q} &= \left(
		\begin{array}{cc}
			H_{\bm k+\bm q}&\bm \Delta_{\bm k}\\
			\bm \Delta^\dagger_{\bm k}&-H_{\bm k-\bm q}
		\end{array}
	\right), \\
	\hat{\Psi}^\dagger_{\bm k,\bm q} &= \left(
	\begin{array}{cc}
		\hat{\bm c}^\dagger_{\bm k+\bm q}&\hat{\bm c}^T_{-\bm k+\bm q}U_{\mathcal{T}}^\dagger
	\end{array}
\right).
\end{align}
%(see Appendix~\ref{sec:ana_derivation} for more details).
Here, we denote
$
\hat{\bm c}^\dagger_{\bm k} = (
	\begin{array}{cccccc}
		\hat{c}^\dagger_{\uparrow 1 \bm k}&\cdots&\hat{c}^\dagger_{\uparrow f \bm k}&\hat{c}^\dagger_{\downarrow 1 \bm k}&\cdots&\hat{c}^\dagger_{\downarrow f \bm k}
	\end{array}
)
$ , where
$\hat{c}^{\dagger}_{\sigma l \bm k}$ is the creation operator for spin $\sigma$ and the other internal degree of freedom $l$.
We consider general cases, including multi-orbital and multi-sublattice systems, and $f$ is the total number of degrees of freedom other than spin. %See Appendix~\ref{sec:ana_derivation} for more details.

The off-diagonal part $\bm \Delta_{\bm k}= \bm \Delta^{\rm g}_{\bm k}+\bm \Delta^{\rm u}_{\bm k}$ is the gap function in the matrix representation, where $\bm \Delta^{\rm g(u)}_{\bm k}$ is the $\mathcal{P}$-even (odd) component of the pair potential.
Coexistence of Cooper pairs with different parities, i.e. parity-mixed state, leads to broken $\mathcal{P}$-symmetry.
Furthermore, the $\mathcal{T}$-symmetry breaking is theoretically predicted~\cite{wang2017topological,yang2020single,sergienko2004mixed}, when the normal state preserves the $\mathcal{P}$-symmetry,
Thus, we assume the $\pm\pi/2$ phase difference between $\bm \Delta^{\rm g}_{\bm k}$ and $\bm \Delta^{\rm u}_{\bm k}$, consistent with the theoretical prediction~\cite{wang2017topological,yang2020single,sergienko2004mixed}.
As a result the $\mathcal{P}$- and $\mathcal{T}$-symmetry are broken by the parity-mixed gap function while the $\mathcal{PT}$-symmetry is preserved.
In addition, to make the anapole moment finite, throughout the paper, we assume the gap function $\bm \Delta_{\bm k}$ belongs to polar irreducible representation. 

%The off-diagonal part $\bm \Delta_{\bm k}= \bm \Delta^{\rm g}_{\bm k}+\bm \Delta^{\rm u}_{\bm k}$ is the gap function in the matrix representation, where $\bm \Delta^{\rm g(u)}_{\bm k}$ is the $\mathcal{P}$-even (odd) component of the pair potential. We assume that the $\mathcal{P}$- and $\mathcal{T}$-symmetry are broken by the parity-mixed gap function while the $\mathcal{PT}$-symmetry is preserved. To make the anapole moment finite, throughout the paper, we assume the gap function $\bm \Delta_{\bm k}$ belongs to polar irreducible representation. 

Expanding the free energy by $\bm q$ as $F_{\bm q} = \bm T\cdot\bm q + \cdots$, we obtain the anapole moment as,
\begin{eqnarray}
	T_{\mu} &=& \dfrac{1}{2}\sum_{\bm k}\sum_af(E_{a\bm k})
	\bra{\psi_{a\bm k}}\partial_\mu H^+_{\bm k}\ket{\psi_{a\bm k}},\label{eq:anapole_moment}\\
	H^+_{\bm k} &=& \left(
		\begin{array}{cc}
			H_{\bm k}&0\\
			0&H_{\bm k}
		\end{array}
	\right).
\end{eqnarray}
Here, we use the eigenvalue equation $H^{\BdG}_{\bm k}\ket{\psi_{a\bm k}}=E_{a\bm k}\ket{\psi_{a\bm k}}$ with $H^{\BdG}_{\bm k} \equiv H^{\BdG}_{{\bm k},{\bm 0}}$ and the Fermi-distribution function $f(E)$.
The derivation of Eq.~\eqref{eq:anapole_moment} is shown in Appendix~\ref{sec:ana_derivation}.
When the anapole moment in superconductors $T_\mu$ is finite, a superconducting state due to Cooper pairs with finite center of mass momenta becomes most stable.

To obtain further insights, using the Bloch wave function which follows $H_{\bm k}\ket{u_{n\chi\bm k}}= \epsilon_{n\bm k}\ket{u_{n\chi\bm k}}$,  we expand $\ket{\psi_a(\bm k)}$ as
$
  \ket{\psi_{a\bm k}} = \left(
  \begin{array}{cc}
    \sum_{n,\chi}\phi_{n\chi\bm k}^{a+}\ket{u_{n\chi\bm k}}&
    \sum_{n,\chi}\phi_{n\chi\bm k}^{a-}\ket{u_{n\chi\bm k}}
  \end{array}
  \right)^T.
$
Because of Kramers degeneracy, we distinguish two degenerate bands by the helicity $\chi = \uparrow\downarrow$.
Here, $\phi_{n\chi\bm k}^{a\pm}$ is the matrix element of the unitary matrix which diagonalizes the band representation of the BdG Hamiltonian.
After calculations, the anapole moment Eq.~\eqref{eq:anapole_moment} is divided into two parts,
\begin{eqnarray}
&&T_{\mu} = T^{\rm velo}_{\mu} + T^{\rm geom}_{\mu},\label{eq:anapole_two_term}
\end{eqnarray}
where
\begin{eqnarray}
T^{\rm velo}_{\mu} &=& \sum_{\bm k}\sum_{n,\chi}C_{n \chi n\chi\bm k}\partial_{\mu}\epsilon_{n\bm k}\label{eq:assym},\label{eq:anapole_velo}\\
T^{\rm geom}_{\mu} &=& \sum_{\bm k}\sum_{n\neq m,\chi\chi^\prime}C_{n\chi m\chi^\prime \bm k}
\notag\\
&\times&
(\epsilon_{m\bm k}-\epsilon_{n\bm k})\braket{u_{n \chi \bm k}\vert\partial_\mu u_{m \chi^\prime\bm k}}\label{eq:geom},\\
C_{n\chi m\chi^\prime \bm k} &=& \dfrac{1}{2}\sum_{a}f(E_{a\bm k})\left(\phi_{n\chi \bm k}^{a+*}\phi_{m\chi^\prime \bm k}^{a+}
+\phi_{n\chi \bm k}^{a-*}\phi_{m \chi^\prime\bm k}^{a-}\right).\notag\\
\end{eqnarray}
$T^{\rm velo}_{\mu}$ in Eq.~\eqref{eq:assym} is called the group velocity term as it contains the group velocity $\partial_\mu\epsilon_{n\bm k}$.
In the next section, using the general two-band model, we show that this term arises from the asymmetric BS.
Equation~\eqref{eq:geom} for $T^{\rm geom}_{\mu}$ is named the geometric term because it contains the Berry connection $\braket{u_{n\chi\bm k}\vert\partial_\mu u_{m\chi^\prime\bm k}}$.
Through the Berry connection in the geometric term, the geometric properties of Bloch electrons may contribute to the anapole moment.
%More specifically, the asymmetric distribution of not only the quasiparticles' number density but also the Berry connection in the momentum space, \YY{both of which can be induced by the $\mathcal{PT}$-symmetric parity-mixed pairing,} may cause the anapole superconductivity.
Some conditions have to be satisfied for a finite group velocity term, which vanishes in simple models~\cite{kanasugi2022anapole}. On the the other hand, the geometic term has been overlooked in the previous study.
Owing to the geometric term, the anapole superconductivity becomes recognized as a ubiquitous feature of the $\mathcal{PT}$-symmetric mixed-parity pairing state in multiband superconductors.

\section{Origin of anapole superconductivity}\label{sec:origin}
\subsection{General discussion\label{sec:origin_general}}
Before demonstrating the anapole superconductivity due to quantum geometry, we discuss the physical origin and the microscopic process of the anapole moment using the Ginzburg-Landau (GL) theory. We also discuss their relation to the group velocity and geometric terms.
Up to the second-order of the gap function $\bm \Delta_{\bm k}$, the anapole moment is given by, 
\begin{eqnarray}
    T^{\rm GL}_{\mu} &&= \dfrac{1}{\beta}\sum_{\bm k\omega_n}{\rm tr}
	\left[\mathcal{G}^{\rm p}_{\bm k\omega_n}\partial_\mu H_{\bm k}\mathcal{G}^{\rm p}_{\bm k\omega_n}\bm \Delta^{\rm g}_{\bm k}\mathcal{G}^{\rm h}_{\bm k\omega_n} \bm \Delta^{{\rm u}\dagger}_{\bm k}\right.\notag\\
	&&-\left.
	\mathcal{G}^{{\rm p}}_{\bm k\omega_n}\partial_\mu H_{\bm k}\mathcal{G}^{\rm p}_{\bm k\omega_n} \bm \Delta^{{\rm u}\dagger}_{\bm k}\mathcal{G}^{\rm h}_{\bm k\omega_n}\bm \Delta^{\rm g}_{\bm k}
	\right]+({\rm g}\leftrightarrow{\rm u})\label{eq:ana_gl},
\end{eqnarray}
where ${\rm tr}$ represents the trace over normal state degrees of freedom.
Here, $\mathcal{G}^{\rm p(h)}_{\bm k\omega_n}=(i\omega_n\mp H_{\bm k})^{-1}$ is the Green function for the particle (hole) part. The derivation of the formula~\eqref{eq:ana_gl} is shown in Appendix~\ref{appendix:ana_gl}.
From this formula, we see that $\mathcal{P}$- and $\mathcal{T}$-symmetry breaking is needed for the anapole superconductivity
(see also Appendix~\ref{appendix:ana_gl}).

We can rewrite the formula in the Bloch band basis,
\begin{align}
    T^{\rm GL}_{\mu} &= \dfrac{1}{\beta}\sum_{\bm k\omega_n}\sum_{nmp}\sum_{\chi_n\chi_m\chi_p}C^{\rm GL}_{nmp{\bm k}\omega_n}
	{\rm tr}\left[P_{n\chi_n \bm k}\partial_{\mu}H_{\bm k}P_{m\chi_m \bm k}\right.\notag\\
	&\times\left.
	\left(
	\bm \Delta^{\rm g}_{\bm k}P_{p\chi_p \bm k} \bm \Delta^{{\rm u}\dagger}_{\bm k}
	-\bm \Delta^{{\rm u}\dagger}_{\bm k}P_{p\chi_p \bm k} \bm \Delta^{\rm g}_{\bm k}
	\right)
	\right]+({\rm g}\leftrightarrow{\rm u})\label{eq:ana_gl_b},
\end{align}
where $C_{nmp\bm k\omega_n}^{\rm GL}=(i\omega_n-\epsilon_{n\bm k})^{-1}(i\omega_n-\epsilon_{m\bm k})^{-1}(i\omega_n+\epsilon_{p\bm k})^{-1}$,
and $P_{n\chi_n \bm k}=\ket{u_{n\chi_n \bm k}}\bra{u_{n\chi_n \bm k}}$ is the projection operator. %of the normal state.
For $n=m=p$, the summand of Eq.~\eqref{eq:ana_gl_b} vanishes (see Appendix~\ref{appendix:degenerate} for details).
Therefore, at least two pairs of $n$, $m$ and $p$ must be nonequivalent for a finite contribution to the anapole moment.
In other words, two interband processes are necessary for the anapole superconductivity.

The above necessary condition for $n$, $m$ and $p$ can be satisfied in three cases.
The first case, $n=m \neq p$, corresponds to the group velocity term, and the odd- and even-parity interband pairings play the role of two interband processes.
In the following subsection, it is shown that the group velocity term is closely related to the asymmetric BS.
The effect of asymmetric BS on the group velocity term is also discussed in Appendix~\ref{appendix:velocity}.

The remaining two cases correspond to the geometric term since the Berry connection is necessary.
%, in which one interband effect is given by the normal state.
In the second case, $n\neq m$ and $n = p$ (or $m=p$), the Berry connection of Bloch electrons and either odd-parity or even-parity interband pairing play the role of two interband processes.
%We note that either even-parity or odd-parity intraband pairing component is also needed in this case.}
Finally, in the third case, $n\neq m\neq p \neq n$, all of the even-parity interband pairing, odd-parity interband pairing, and the Berry connection appear in the contribution to the anapole moment.
In both cases, via the Berry connection, the Bloch electrons undergo an interband transition from the initial band to the different band, which is coupled to the initial band through the interband Cooper pairs. Thus, the two or more interband processes, due to the Berry connection and interband Cooper pairs, induce the anapole moment. which is a physical picture of quantum-geometry-induced anapole superconductivity.

\subsection{General two-band model}\label{sec:two-band_model}
Next, for a more transparent understanding, we derive the anapole moment
%the necessary conditions for the anapole superconductivity 
in general two-band superconductors with Kramers degeneracy. Although we here adopt a two-band model, the following results can be applied to any multi-band model with multiple bands near the Fermi surface.
The normal-state Hamiltonian is written as, 
\begin{eqnarray}
H_{\bm k} = h_{0\bm k}\bm 1+\bm h_{\bm k}\cdot \bm \gamma\label{eq:gamma_hk},
\end{eqnarray}
by using the $4\times4$ gamma matrices $\bm \gamma = (\begin{array}{ccc}
    \gamma_1& \cdots & \gamma_5 
\end{array})$
that anti-commute each other.
Here, $h_{0\bm k}$ and $\bm h_{\bm k} = (\begin{array}{ccc}
    h_{1\bm k} & \cdots & h_{5\bm k}  
\end{array})$
depend on the details of the model.
The energy dispersion is given by $\epsilon_{\pm,\bm k} =  h_{0{\bm k}}\pm\vert{\bm h_{\bm k}}\vert$.
Note that the 4 by 4 normal-state Hamiltonian has two bands due to the Kramers degeneracy.

The $\mathcal{PT}$-symmetric parity-mixed pair potential is expressed as~\cite{kanasugi2022anapole},
\begin{eqnarray}
&\bm \Delta_{\bm k} = \bm \Delta^{\rm g}_{\bm k}+\bm \Delta^{\rm u}_{\bm k},\\
&\bm \Delta^{\rm g}_{\bm k} = \eta_{0 \bm k}\bm 1+\bm \eta_{\bm k}\cdot\bm \gamma,\,\,\ \bm \Delta^{\rm u}_{\bm k}=\frac{i}{2}\sum_{ij}\tilde{\eta}_{ij\bm k}\gamma_i\gamma_j,
\end{eqnarray}
where $\eta_{0 \bm k}$, $\bm \eta_{\bm k} = (\begin{array}{ccc}
    \eta_{1\bm k} & \cdots & \eta_{5\bm k}  
\end{array})$ and $\tilde{\eta}_{ij \bm k} = -\tilde{\eta}_{ji \bm k}$ are the complex valued order parameters for even- and odd-parity pairing channels. 
Here, taking appropriate U(1) gauge, $\eta_{0\bm k}$ and $\eta_{i\bm k}$ are real while $\tilde{\eta}_{ij\bm k}$ becomes pure imaginary.

%The relations, $\tilde{\eta}_{ij\bm k}/\eta_{0 \bm k}= e^{\pm i\pi/2}\vert\tilde{\eta}_{ij\bm k}\vert/\vert\eta_{0 \bm k}\vert$ and $\tilde{\eta}_{ij\bm k}/\eta_{i \bm k}= e^{\pm i\pi/2}\vert\tilde{\eta}_{ij\bm k}\vert/\vert\eta_{i \bm k}\vert$, are satisfied to preserve the $\mathcal{PT}$-symmetry with breaking the $\mathcal{T}$-symmetry.
%\SK{The relation $\mathrm{arg}(\eta_{i\bm k}/\eta_{j \bm k})= \{0,\pi\}$, $\mathrm{arg}(\tilde{\eta}_{ij\bm k}/\tilde{\eta}_{i'j'\bm k})= \{0,\pi\}$, and $\mathrm{arg}(\tilde{\eta}_{ij\bm k}/\eta_{0 \bm k})= \pm\pi/2$ are satisfied to preserve the $\mathcal{PT}$-symmetry with breaking the $\mathcal{T}$-symmetry.}

Because of the Kramers degeneracy, the particle Green function can be projected to the two degenerate bands as~\cite{denys2021origin}
\begin{eqnarray}
&&\mathcal{G}^{\rm p}_{\bm k\omega_n}
	=a_{\bm k\omega_n} {\bm 1}+b_{\bm k\omega_n}\tilde{H}_{\bm k},\\
	&&a_{\bm k\omega_n}=\dfrac{1}{2}\sum_{\pm}(i\omega_n-h_{0\bm k}\pm \vert\bm h_{\bm k}\vert)^{-1},\\
	&&b_{\bm k\omega_n}=\dfrac{1}{2}\sum_{\pm}\mp(i\omega_n-h_{0\bm k}\pm \vert\bm h_{\bm k}\vert)^{-1},
\end{eqnarray}
with $\tilde{H}_{\bm k}=(\bm h_{\bm k}\cdot\bm \gamma)/\vert\bm h_{\bm k}\vert=\hat{\bm h}_{\bm k}\cdot\bm \gamma$.
The hole Green function is also given by,
%$
\begin{eqnarray}
\mathcal{G}^{\rm h}_{\bm k\omega_n}
=c_{\bm k\omega_n} {\bm 1} + d_{\bm k\omega_n} \tilde{H}_{\bm k},
%=a_{\bm k-\omega_n} {\bm 1} + b_{\bm k-\omega_n}\tilde{H}_{\bm k}.
\end{eqnarray}
%$
where $a_{\bm k\omega_n}=-c_{\bm k-\omega_n}$ and 
$b_{\bm k\omega_n}=-d_{\bm k-\omega_n}$.
Hereafter, we omit the $(\bm k, \omega_n)$ dependence for simplicity.
Inserting these expressions of Green functions into Eq.~\eqref{eq:ana_gl}, we get
\begin{eqnarray}
	&&T_{\mu}^{\rm GL} = \dfrac{1}{\beta}\sum_{\bm k}\sum_{\omega_n}
	\left(a^2c{\rm tr}\left[\partial HM^{(1)}_-\right]+a^2d{\rm tr}\left[\partial HM^{(2)}_-\right]\right.\notag\\
	&&
	+abc{\rm tr}\left[\left\{\partial H,\tilde{H}\right\}M^{(1)}_-\right]
	+abd{\rm tr}\left[\left\{\partial H,\tilde{H}\right\}M^{(2)}_-\right]
	\notag\\
	&&+b^2c{\rm tr}\left[\tilde{H}\partial H\tilde{H}M^{(1)}_-\right]+\left.b^2d{\rm tr}\left[\tilde{H}\partial H\tilde{H}M^{(2)}_-\right]
	\right)\label{eq:ana_gl4}.
\end{eqnarray}
Here, we introduce the $\mathcal{P}$- and $\mathcal{T}$-odd bilinear products~\cite{brydon2018bogoliubov,brydon2019loop,denys2021origin,kanasugi2022anapole},
\begin{align}
M^{(1)}_- &= \left[\bm \Delta^{\rm g},\bm \Delta^{{\rm u}\dagger}\right]+({\rm g}\leftrightarrow{\rm u}),\\
M^{(2)}_- &= \left[\bm \Delta^{\rm g}\tilde{H}\bm \Delta^{{\rm u}\dagger}-\bm \Delta^{{\rm u}\dagger}\tilde{H}\bm \Delta^{\rm g}\right]+({\rm g}\leftrightarrow{\rm u}).
\end{align}
According to Eq.~\eqref{eq:ana_gl4}, the presence of finite bilinear products is a necessary condition for the anapole superconductivity.

One of the bilinear products $M^{(1)}_-$ %corresponds to the first condition of the asymmetric BS~\cite{kanasugi2022anapole} and written by,
is obtained as,
\begin{align}
     M^{(1)}_- &= \bm m_1\cdot\bm \gamma,\\
     [{\bm m}_1]_j &= -4 \sum_{i(\neq j)}{\rm Im}\left[\eta_{i}\tilde{\eta}^*_{ij}\right].
\end{align} 
It has been shown that $M^{(1)}_-$ is needed for the asymmetric BS~\cite{kanasugi2022anapole}, and thus, ${\bm m}_1$ represents the role of the asymmetric BS. 
More specifically, the necessary condition of the asymmetric BS is given by $\bm m_1\cdot\hat{\bm h}\neq0$~\cite{kanasugi2022anapole}.
To elucidate the origin and physical meaning of another bilinear product $M^{(2)}_{-}$, we introduce the interband and intraband superconducting fitness (SCF)~\cite{ramires2016identifying,ramires2018tailoring},  $F^{\rm C}_{\rm g(u)}$ and $F^{\rm A}_{\rm g(u)}$, which are defined by 
\begin{eqnarray}
    &F^{\rm C}_{\rm g(u)} = \left[\tilde{H},\bm \Delta^{\rm g(u)}\right], \ F^{\rm A}_{\rm g(u)} = \left\{\tilde{H},\bm \Delta^{\rm g(u)}\right\}.
\end{eqnarray}
Using this, we can rewrite $M^{(2)}_-$ as,
\begin{align}
    M^{(2)}_-&=\dfrac{1}{4}\left(
    \left[F_{\rm g}^{\rm A},\bm \Delta^{{\rm u}\dagger}\right]
    +\left[F_{\rm u}^{\rm A},\bm \Delta^{{\rm g}\dagger}\right]
    \right.\notag\\
    &\left.
    -\left\{F_{\rm g}^{\rm C},\bm \Delta^{{\rm u}\dagger}\right\}-\left\{F_{\rm u}^{\rm C},\bm \Delta^{{\rm g}\dagger}\right\}
    \right)+{\rm h.c.}\label{eq:m2_scf}
\end{align}
This means that both interband and intraband pairings lead to a finite bilinear product $M^{(2)}_{-}$.
Each term in Eq.~\eqref{eq:m2_scf} is calculated as,
\begin{eqnarray}
   \left[F_{\rm g}^{\rm A},\bm \Delta^{{\rm u}\dagger}\right]+{\rm h.c.} &=& 2\bm m_2\cdot\bm\gamma,\\
   \left[F_{\rm u}^{\rm A},\bm \Delta^{{\rm g}\dagger}\right]+{\rm h.c.} &=& 2\bm m_3\cdot\bm\gamma,\\
   \left\{F_{\rm g}^{\rm C},\bm \Delta^{{\rm u}\dagger}\right\}+{\rm h.c.} &=& -2\bm m_3\cdot\bm\gamma +2(\bm m_1\cdot\hat{\bm h} )\bm 1,\\
   \left\{F_{\rm u}^{\rm C},\bm \Delta^{{\rm g}\dagger}\right\}+{\rm h.c.} &=& -2\bm m_2\cdot\bm\gamma +2( \bm m_1\cdot\hat{\bm h})\bm 1,
\end{eqnarray}
where,
\begin{eqnarray}
    &&\left[\bm m_2\right]_j =-4\sum_{i(\neq j)}\hat{h}_i{\rm Im}\left[\eta_{0}\tilde{\eta}^*_{ij}\right]\label{eq:m2_two_band},\\
    &&\left[\bm m_3\right]_j =-2\sum_{i_1 i_2 i_3  i_4}\varepsilon_{i_1i_2i_3i_4j}\hat{h}_{i_1}{\rm Im}\left[\eta_{i_2}\tilde{\eta}^*_{i_3i_4}\right]\label{eq:m3_two_band}.
\end{eqnarray}
Here, we use the relationship,
\begin{eqnarray}
   \gamma_j = \dfrac{-1}{4!}\sum_{i_1 i_2 i_3 i_4}\epsilon_{ji_1i_2i_3i_4}\gamma_{i_1}\gamma_{i_2}\gamma_{i_3}\gamma_{i_4},
\end{eqnarray}
with the Levi-Civita tensor $\epsilon_{i_1i_2i_3i_4i_5}$.
Inserting these expressions into Eq.~\eqref{eq:m2_scf}, we obtain,
\begin{eqnarray}
    &&M^{(2)}_-=-(\bm m_1\cdot\hat{\bm h})\bm 1+(\bm m_2+\bm m_3)\cdot\bm \gamma.
\end{eqnarray}
The first term comes from the asymmetric BS, which originates from the interband SCF.
On the other hand, the second and third terms arise from either the intraband SCF and interband SCF. To be more precise, the term $\bm m_{2}\cdot\bm \gamma$ ($\bm m_{3}\cdot\bm \gamma$) needs the  $\mathcal{P}$-even ($\mathcal{P}$-odd) intraband SCF or the  $\mathcal{P}$-odd  ($\mathcal{P}$-even) interband SCF.
%\TK{Note that the intraband SCF does not ensure the presence of the intraband pairing, mentioned in the previous subsection, because our model has the Kramers degeneracy.
%However, it is an important point that the intraband SCF can be finite even in the absence of the interband pairing.}
This implies that $\bm m_{2}$ ($\bm m_{3}$) contains information of even-parity (odd-parity) intraband pairing and odd-parity (even-parity) interband pairing.
Thus, introducing the SCF helps understand the role of $\mathcal{P}$-even and $\mathcal{P}$-odd interband parings in the anapole superconductivity.

After a tedious but straightforward calculation we can get the group velocity term and geometric term in the GL theory, 
\begin{eqnarray}
T^{\rm GL} &=& T^{\rm GL:velo} + T^{\rm GL:geom}, \\
    T^{\rm GL:velo} &=& \dfrac{4}{\beta}\sum_{\bm k\omega_n}\left[\left(2abc-a^2d-b^2d\right)\partial h_0\right.\notag\\
    &&-\left.\left(2abd-a^2c-b^2c\right)\partial \vert \bm h\vert \right]\bm m_1\cdot\hat{\bm h},\\
    T^{\rm GL:geom} &=& \dfrac{4}{\beta}\sum_{\bm k\omega_n}(a^2-b^2)\vert\bm h\vert\partial \hat{\bm h}\cdot\left[c\bm m_1\right.%\notag\\
    \left.+d\left(\bm m_2+\bm m_3\right) \right].\notag\\\label{eq:geom_orign}
\end{eqnarray}
%\YYS{The group velocity term is induced only by the asymmetric BS, $\bm m_1\cdot \hat{\bm h}$.}
It should be noticed that the group velocity term contains a factor $\bm m_1\cdot \hat{\bm h}$, which is closely related to the asymmetric BS.
%This contribution corresponds to the case of $n=m \neq p$ in Eq.~\eqref{eq:ana_gl_b}.
Therefore, we conclude that the asymmetric BS is an origin of the group velocity term (see also Appendix~\ref{appendix:velocity} for a simple case).

On the other hand, the Berry connection is essential for the geometric term, as Eq.~\eqref{eq:geom_orign} contains $\partial\hat{\bm h}$ which makes the Berry connection finite.
More specifically, the geometric term arises from various contributions, which are understood by Eq.~\eqref{eq:geom_orign}. 
For the frist term of Eq.~\eqref{eq:geom_orign}, in addition to the Berry connection, asymmetric BS is also essential in this contribution.
In contrast, the second (third) term of Eq.~\eqref{eq:ana_gl_b} can be finite without even- (odd)-parity interband pairing.
Thus, either odd-parity or even-parity interband pairing causes the anapole superconductivity owing to the quantum geometry, although the group velocity term needs both odd-parity and even-parity interband pairing. 
In other words, whenever the group velocity term is finite, finite Berry connection ensures the presence of the geometric term. Moreover, even when the group velocity term is absent, the geometric term can be finite due to $m_2$ and $m_1$.
Therefore, necessary conditions for the anapole superconductivity are relaxed by appropriately considering the quantum geometric effect, which was neglected in Ref.~\onlinecite{kanasugi2022anapole}. 
Later, we will also show that the anapole moment is dominated by the geometric term at low temperatures.

%On the other hand, the Berry connection is essential for the geometric term, as Eq.~\eqref{eq:geom_orign} contains $\partial\hat{\bm h}$ which makes the Berry connection finite.
%More specifically, the geometric term arises from various processes and the contributions from each process are understood by Eq.~\eqref{eq:geom_orign}. 
%The case of $n\neq m \neq p \neq n$ in Eq.~\eqref{eq:ana_gl_b} corresponds to the first term of Eq.~\eqref{eq:geom_orign}.
%In addition to the Berry connection, asymmetric BS is also essential in this contribution. 
%The case of $n=p$ or $m=p$ with $n\neq m$ in Eq.~\eqref{eq:ana_gl_b} gives the second and third terms of Eq.~\eqref{eq:geom_orign}, consistent with the fact that both interband and intraband pairing are necessary for $\bm m_2$ and $\bm m_3$.
%Thus, either odd-parity or even-parity interband pairing causes the anapole superconductivity owing to the quantum geometry, although the group velocity term needs both odd-parity and even-parity interband pairing. Therefore, necessary conditions for the anapole superconductivity are relaxed by appropriately considering the quantum geometric effect, which was neglected in Ref.~\onlinecite{kanasugi2022anapole}. 
%Later, we will also show that the anapole moment is dominated by the geometric term at low temperatures.

\section{Quantum-geometry-induced anapole superconductivity in UTe$_2$}

In this section, we first show a general theory for anapole superconductivity in locally noncentrosymmetric superconductors (Sec.~\ref{sec:orign_locally}) and next focus on UTe$_2$ (Secs.~\ref{sec:symmetry_analysis} and \ref{sec:UTe2}).

\subsection{Locally noncentrosymmetric superconductors\label{sec:orign_locally}}

Many exotic superconductors of recent interest have multiple sublattices which do not lie on the inversion center.  UTe$_2$~\cite{Aoki_review2022} and CeRh$_2$As$_2$~\cite{khim2021field} are examples of such locally noncentrosymmetric superconductors, and a part of materials is listed in a review article~\cite{fischer2022superconductivity}.
Before demonstrating the quantum-geometry-induced anapole superconductivity in UTe$_2$, we show that the locally noncentrosymmetric superconductors are generically the platform of anapole superconductivity and clarify the conditions for it. %based on the GL expansion. %a minimal model for UTe$_2$. 
While the following discussions in this subsection are based on the GL expansion, %this can give a good prediction about the behavior of 
the BCS theory reproduces the results for the anapole moment quantitatively, %in the low-temperature region 
as shown in Sec.~\ref{sec:UTe2}.

We consider the locally noncentrosymmetric two-sublattice model~\cite{kanasugi2022anapole}, which is adopted as a minimal model for UTe$_2$ later,
\begin{eqnarray}
    &&H = \xi \sigma_0\otimes\tau_0 + w_x\sigma_0\otimes\tau_x+w_y\sigma_0\otimes\tau_y+\bm g\cdot\bm \sigma\otimes\tau_z\notag,\\
    \label{eq:hamiltonian_ute2}\\
    &&\bm \Delta^{\rm g} = \Delta^{\rm g}\left[\,\sum_{\mu=0,x,y}\phi_{\rm g}^{\mu}\sigma_0\otimes\tau_\mu+\bm d_{\rm g}^z\cdot\bm \sigma\otimes\tau_z\right],\\
    &&\bm \Delta^{\rm u} = \Delta^{\rm u}\left[\,\sum_{\mu=0,x,y}\bm d_{\rm u}^{\mu}\cdot\bm \sigma\otimes\tau_\mu+\phi_{\rm u}^z \sigma_0\otimes\tau_z\right].
\end{eqnarray}
Here, $\sigma_\mu$ and $\tau_\mu$ are the Pauli matrices for the spin and sublattice spaces, %which arises from the ladder structure of two U atoms, respectively.
$\xi$ is the single-particle kinetic energy, and 
$\bm g = (g_x, g_y, g_z)
%\begin{array}{ccc}
%    g_x & g_y & g_z
%\end{array})
$ is the staggered-type ASOC due to the local $\mathcal{P}$-symmetry breaking at atomic sites.
%antisymmetric spin-orbit coupling (ASOC) 
For example, in UTe$_2$, U atoms form a ladder structure, which consists of two sublattices lacking the $\mathcal{P}$-symmetry at the atomic sites. 
The local point group descends to $C_{2v}$ from $D_{2h}$, and therefore, the Rashba ASOC naturally appears. %at the U sites. %due to the local $\mathcal{P}$-symmetry breaking. %which arises from the ladder structure of U atoms, 
Since the two sublattices are related by the global $\mathcal{P}$-symmetry, the Rashba ASOC shows a staggered form proportional to $\tau_z$.
%On the two sublattices of U ions, the opposite coupling constant $\pm\alpha$ of Rashba ASOC between two U atoms is naturally expected.

%For the superconducting order parameter,
The superconducting pair potentials are divided into the 
spin-singlet component $\phi_{\rm g(u)}^\mu$ and the spin-triplet component $\bm d_{\rm g(u)}^\mu$. %are introduced.
%, which satisfies the relationships, $\phi_{\bm k}^{0(x)}=\phi_{-\bm k}^{0(x)}$, $\phi_{\bm k}^{y}=-\phi_{-\bm k}^{y}$, and $\bm d_{\bm k}^\mu=-\bm d_{-\bm k}^\mu$.
%$\bm \Delta_{\rm g(u)}$が偶(奇)パリティであるので，波数等の詳細は明記しなくても良いかと思いました
The local inversion symmetry breaking also leads to sublattice-dependent parity-mixing of the pair potential.
Thus, the sublattice-independent spin-singlet (spin-triplet) pairing component and the staggered spin-triplet (spin-singlet) one coexist in the even-parity (odd-parity) pair potential. 
%\YYS{Since the anapole superconductivity needs that the pair potential belongs to the polar irreducible representation, we assume $\bm \Delta_{\rm g}$ and $\bm \Delta_{\rm u}$ belong to the $A_{\rm g}$ and $B_{\rm 1(2,3)u}$ irreducible representation, respectively, which are proposed in the recent theoretical study~\cite{ishizuka2021periodic}.}
%In addition, 
To preserve the $\mathcal{PT}$-symmetry while breaking the $\mathcal{T}$-symmetry, the relative phase between the complex-valued order parameters $\Delta^{\rm g}$ and $\Delta^{\rm u}$ is assumed to be $\pi/2$, and thus, $4{\rm Im}(\Delta^{\rm g}\Delta^{\rm u*}) \ne 0$.

\begin{table}[tbp]
 \caption{Correspondence between Pauli and Dirac matrices.}
 \label{table:pauli_dirac}
 \centering
  \begin{tabular}{cc}
   \hline
   Pauli & Dirac\\
   \hline \hline
   $\bm \sigma\otimes \tau_z$ & $(\begin{array}{ccc}
        \gamma_1&\gamma_2&\gamma_3 \\
   \end{array})$ \\
    $\bm \sigma\otimes \tau_y$ & $(\begin{array}{ccc}
        -i\gamma_1\gamma_4&-i\gamma_2\gamma_4&-i\gamma_3\gamma_4 \\
   \end{array})$ \\
    $\bm \sigma\otimes \tau_x$ & $(\begin{array}{ccc}
        i\gamma_1\gamma_5&i\gamma_2\gamma_5&i\gamma_3\gamma_5 \\
   \end{array})$ \\
    $\bm \sigma\otimes \tau_0$ & $(\begin{array}{ccc}
        -i\gamma_2\gamma_3&i\gamma_1\gamma_3&-i\gamma_1\gamma_2 \\
   \end{array})$ \\
    $\sigma_0\otimes \bm \tau$ & $(\begin{array}{ccc}
        \gamma_4&\gamma_5&-i\gamma_4\gamma_5 \\
   \end{array})$ \\
   \hline
  \end{tabular}
\end{table}

We show the correspondence between the Pauli matrices and the Dirac matrices in Table~\ref{table:pauli_dirac}~\cite{paulicomment},
from which the condition for anapole superconductivity can be derived based on the discussions in Sec.~\ref{sec:origin}. The results are summarized in Table~\ref{table:ana_UTe2_condition}, where conditions for the finite group velocity term and the geometric term are explicitly presented.
Here, we define $\hat{\bm g} = {\bm g}/{\vert \bm h\vert}$ and $\hat{w}_{x(y)} = {w_{x(y)}}/{\vert \bm h\vert}$ with $\vert \bm h\vert = \sqrt{w_x^2+w_y^2+\vert\bm g\vert^2}$.
%\TK{We note that we omit the coefficient $4{\rm Im}(\Delta^{\rm g}\Delta^{\rm u*})$ for the Condition (Pauli).}
Note that all the terms of anapole moment is proportional to  $4{\rm Im}(\Delta^{\rm g}\Delta^{\rm u*})$, which is finite by assumption.
The geometric term can be finite in a relatively simple situation. 
For example, the ${\bm m}_2$-term gives a finite anapole moment when $\phi_{\rm g}^0 (\hat{\bm g}\times\bm d_{\rm u}^0)\cdot\partial \hat{\bm g}\neq0$. This condition is satisfied in the presence of an usual spin-singlet pairing component $\phi_{\rm g}^0$ and an interband pairing component $\hat{\bm g}\times\bm d_{\rm u}^0$ %induces the anapole superconductivity 
when the corresponding Berry connection is finite.
If the $s+ip$-wave  pairing state is realized in UTe$_2$ as proposed~\cite{ishizuka2021periodic}, the even-parity $s$-wave component, $\phi_{\rm g}^0$, and the odd-parity $p$-wave component, $\bm d_{\rm u}^0$, naturally exist, and the Berry connection arises from the staggered Rashba ASOC. %may have to exist in UTe$_2$.
Thus, the anapole superconductivity %is considered to be easily realized 
is likely to occur in UTe$_2$ owing to the quantum geometry when the $s+ip$-wave state is stabilized.
% appears in UTe$_2$.

\begin{table}[htbp]
 \caption{Conditions for the anapole superconductivity in locally noncentrosymmetric systems. The corresponding SCF is also shown.
 We assume $4{\rm Im} (\Delta^{\rm g}\Delta^{\rm u*}) \ne 0$, which is satisfied in the parity-mixed $\mathcal{T}$-symmetry breaking superconductors.
% \TK{We note that we omit the coefficient $4{\rm Im}(\Delta^{\rm g}\Delta^{\rm u*})$ for the Condition (Pauli).}
 When an inequality listed in the table is satisfied, (a) the group velocity term and (b) the geometric term are finite.}
 \label{table:ana_UTe2_condition}
 \centering
  (a) Group velocity term\\
  \begin{tabular}{ccc}
   \hline \hline
    Condition (Dirac)& Condition (Pauli) & SCF \\
   \hline 
     $\partial\epsilon_{\pm}\bm m_1\cdot\hat{\bm h}\neq0$ &
     \begin{tabular}{c}
        $\partial\epsilon_{\pm}\bm d_{\rm g}^z\cdot\bm d_{\rm u}^{y(x)}\hat{w}_{x(y)}\neq0$  \\
        $\partial\epsilon_{\pm}\phi_{\rm g}^{y(x)}\phi_{\rm u}^z\hat{w}_{x(y)}\neq0$    \\
        $\partial\epsilon_{\pm}\phi^{y(x)}_{\rm g}\bm d_{\rm u}^{x(y)}\cdot\hat{\bm g}\neq0$    \\
        $\partial\epsilon_{\pm}\left(\bm d_{\rm g}^z\times\bm d_{\rm u}^0\right)\cdot\hat{\bm g}\neq0$  
     \end{tabular}&
    $F^{C}_{\rm g},F^{C}_{\rm u}\neq0$\\
    \hline \hline\\
  \end{tabular}\\
    (b) Geometric term\\
 \begin{tabular}{ccc}
   \hline \hline
    Condition (Dirac)& Condition (Pauli) & SCF \\
   \hline 
     $\bm m_1\cdot\partial\hat{\bm h}\neq0$ &
     \begin{tabular}{c}
        $\bm d_{\rm g}^z\cdot\bm d_{\rm u}^{y(x)}\partial\hat{w}_{x(y)}\neq0$  \\
        $\phi_{\rm g}^{y(x)}\phi_{\rm u}^z\partial\hat{w}_{x(y)}\neq0$    \\
        $\phi^{y(x)}_{\rm g}\bm d_{\rm u}^{x(y)}\cdot\partial\hat{\bm g}\neq0$    \\
        $\left(\bm d_{\rm g}^z\times\bm d_{\rm u}^0\right)\cdot\partial\hat{\bm g}\neq0$  
     \end{tabular}&
    $F^{C}_{\rm g},F^{C}_{\rm u}\neq0$\\
    \hline
    $\bm m_2\cdot\partial\hat{\bm h}\neq0$ &
    \begin{tabular}{c}
        $\phi_{\rm g}^0\hat{\bm g}\cdot\bm d_{\rm u}^{y(x)}\partial\hat{w}_{x(y)}\neq0$  \\
        $\phi_{\rm g}^0\hat{w}_{y(x)}\phi_{\rm u}^z\partial\hat{w}_{x(y)}\neq0$    \\
        $\phi_{\rm g}^0\hat{w}_{y(x)}\bm d_{\rm u}^{x(y)}\cdot\partial\hat{\bm g}\neq0$    \\
        $\phi_{\rm g}^0\left(\hat{\bm g}\times\bm d_{\rm u}^0\right)\cdot\partial\hat{\bm g}\neq0$  
    \end{tabular}&
    $F^{A}_{\rm g},F^{C}_{\rm u}\neq0$\\
    \hline
    $\bm m_3\cdot\partial\hat{\bm h}\neq0$ &
    \begin{tabular}{c}
        $\hat{\bm g}\cdot(\bm d_{\rm g}^z\times \bm d_{\rm u}^{y(x)})\partial\hat{w}_{y(x)}\neq0$  \\
        $\hat{w}_{x(y)}\bm d_{\rm g}^z\cdot\bm d_{\rm u}^0\partial\hat{w}_{y(x)}\neq0$    \\
        $\hat{\bm g}\cdot(\phi_{\rm g}^{x(y)} \bm d_{\rm u}^0)\partial\hat{w}_{y(x)}\neq0$    \\
        $\hat{w}_{x(y)}(\bm d_{\rm g}^z\times\bm d_{\rm u}^{x(y)})\cdot\partial\hat{\bm g}\neq0$  \\
        $\hat{w}_{x(y)}\phi_{\rm g}^{y(x)}\bm d_{\rm u}^0\cdot\partial\hat{\bm g}\neq0$\\
        $(\hat{\bm g}\times(\phi_{\rm g}^{x(y)}\bm d_{\rm u}^{x(y)}))\cdot\partial\hat{\bm g}\neq0$\\
        $(\hat{\bm g}\times\bm d_{\rm g}^{z}\phi_{\rm u}^z)\cdot\partial\hat{\bm g}\neq0$
    \end{tabular}&
    $F^{C}_{\rm g},F^{A}_{\rm u}\neq0$\\
    \hline
    \hline\\
  \end{tabular}\\    
\end{table}

\subsection{Symmetry classification in the $D_{2h}$ point group}\label{sec:symmetry_analysis}

Here, we show the classification of parity-mixed superconducting states assuming the crystals of $D_{2h}$ point group symmetry with UTe$_2$ in mind.
Combination of the four even-parity irreducible representations ($A_{\rm g}$, $B_{\rm 1g}$, $B_{\rm 2g}$, $B_{\rm 3g}$) and odd-parity ones ($A_{\rm u}$, $B_{\rm 1u}$, $B_{\rm 2u}$, $B_{\rm 3u}$) gives $4 \times 4 =16$ classes of parity-mixed pairing states. They are classified into either the anapole superconductivity or monopole superconductivity. 

The anapole superconducting state has the polarity and shares the symmetry with the magnetic toroidal ordered state, while the monopole superconducting state is an superconducting analog of the magnetic monopole state~\cite{watanabe2018}. 
In the D$_{2h}$ point group, the order parameter of anapole superconductivity breaks the C$_2$ rotation symmetry which flips the polar axis.
%except for one particular direction. 
In contrast, the C$_2$ rotation symmetry of all directions are preserved in the monopole superconducting state, which means that the anapole moment vanishes and the finite-$\bm q$ pairing states are prohibited.

\begin{table}[htbp]
 \caption{Classification of parity-mixed superconducting states in the $D_{2h}$ point group symmetry. Direction of anapole moment is shown for the anapole superconducting state. The other pairing states are monopole superconducting states and represented as "monopole".}
 \label{table:symmetry_analysis}
 \centering
  \begin{tabular}{c|cccc}
& $A_{\rm g}$ & $B_{\rm 1g}$ & $B_{\rm 2g}$ & $B_{\rm 3g}$ \\ \hline
$A_{\rm u}$  & monopole & $T_z$ & $T_y$ & $T_x$ \\
$B_{\rm 1u}$ & $T_z$ & monopole & $T_x$ & $T_y$ \\
$B_{\rm 2u}$ & $T_y$ & $T_x$ & monopole & $T_z$ \\
$B_{\rm 3u}$ & $T_x$ & $T_y$ & $T_z$ & monopole \\
  \end{tabular}
\end{table}

Table~\ref{table:symmetry_analysis} shows the finite component of the anapole moment for the 12 anapole superconducting states, while we denote "monopole" for the monopole superconducting state. For instance, $A_{{\rm g}}+iB_{3{\rm u}}$ pairing state is a anapole superconducting state with anapole moment along the $x$-axis. On the other hand, $A_{{\rm g}}+iA_{{\rm u}}$ pairing state is a non-polar monopole superconducting state, where the anapole moment vanishes.

\subsection{Anapole superconductivity in UTe$_2$}\label{sec:UTe2}
In the theoretical calculation which constructs a 24-orbital model for UTe$_2$~\cite{ishizuka2021periodic}, competing ferromagnetic and antiferromagnetic fluctuations have been shown, implying the competition between the $s$-wave and $p$-wave pairings. Comparing the theoretical results with the experimentally observed multiple superconducting phases, the parity-mixed $\mathcal{T}$-symmetry-broken $s+ip$-wave state was proposed for UTe$_2$. Such $s+ip$-wave superconducting state was also discussed in experimental studies~\cite{rosuel2022thermodynamic,sakai2022-field}. In this scenario, $\mathcal{P}$- and $\mathcal{T}$-symmetry breaking necessary for the anapole superconductivity is predicted. To examine the possible anapole superconductivity in UTe$_2$, we adopt a model for UTe$_2$ and show unique features which have microscopic origins beyond the GL theory.

Here, one of the purposes is to derive the minimal condition and universal properties of anapole superconductivity, which are independent of the detailed band structure. Therefore, we focus on the sublattice and spin degrees of freedom in UTe$_2$, giving rise to multiple bands near the Fermi level. Actually, most of the following results are independent of the details of band structure, as is also discussed in Sec.~\ref{sec:discussion}.
%While our model may not capture the detailed structure of Fermi surfaces, we believe that almost results are independent on the details of band structure.
%To see this, we set the normal-state Hamiltonian Eq.~\eqref{eq:hamiltonian_ute2} as 
Thus, we set the normal-state Hamiltonian Eq.~\eqref{eq:hamiltonian_ute2} as
\begin{eqnarray}
    &w_{x} = w_{y} = 0,\\
    &\xi = t\sum_\mu\cos k_\mu-\mu,\\
    &\bm g = \alpha \left(\sin k_y, -\sin k_x, 0 \right),
    %(\begin{array}{ccc}
%    \sin k_y &-\sin k_x  & 0
%    \end{array}),
\end{eqnarray}
with $(t, \mu, \alpha)=(0.2, 0.4, \pm 0.04)$.
In this model, the energy dispersion is given by $\epsilon_{\pm} =  \xi\pm\vert\bm g\vert$.
%$
%    (\begin{array}{ccc}
%    t &\mu & \alpha \\
%    \end{array}) =
%    (
%    \begin{array}{ccc}
%     0.2  & 0.4 & \pm 0.04 \\
%    \end{array}
%    )
%$.
In Table~\ref{table:ana_UTe2_condition}, some terms of the anapole moment are first order in $\bm g$, implying that the anapole moment may depend on the sign of the ASOC coupling constant $\alpha$. Thus, we examine the two cases, $\alpha= \pm 0.04$. We will actually show that the properties of anapole superconductivity depend on the sign of $\alpha$.

First, we consider the superconducting pair potential,
\begin{align}
    &\bm \Delta^{\rm g}(T) = \Delta_{0}(T)\phi_{\rm g}^{0}\sigma_0\otimes\tau_0,\label{eq:phig}\\
    &\bm \Delta^{\rm u}(T) = i\Delta_{0}(T)d_{{\rm u},z}^{0}\sigma_z\otimes\tau_0, \label{eq:duz}
\end{align}
with $\phi_{\rm g}^0=1$ and $d_{{\rm u},z}^0= \sin k_y$, which belongs to the $A_{{\rm g}}+iB_{3{\rm u}}$ irreducible representation of $D_{2h}$ point group.
The temperature dependence is assumed to be $\Delta_0(T) = \Delta_0^{\rm max}\tanh(1.74\sqrt{(T_{\rm c}-T)/T})$,
% which reproduces the temperature dependence of the gap function,
where $\Delta^{\rm max}_0 = \frac{3.53}{2}T_{\rm c} = 0.02$.
%We note that we assume that transition temperature for even

Based on the GL theory, we expect that the group velocity term vanishes in this case, i.e. $\bm m_1\cdot\hat{\bm h} = 0$. 
However, since the $\bm d$ vector for the B$_{3u}$ representation is not parallel to the $\bm g$ vector for the ASOC of C$_{2v}$ point group, odd-parity interband pairing is always finite in addition to intraband pairing. In other words, $\hat{\bm g}\times\bm d_{\rm u}^0\neq 0$ leads to odd-parity interband pairing.
Therefore, the geometric term is finite because the condition $\bm m_2\cdot\partial\hat{\bm h}=\phi_{\rm g}^0\left(\hat{\bm g}\times\bm d_{\rm u}^0\right)\cdot\partial\hat{\bm g} \ne 0$ in Table~\ref{table:ana_UTe2_condition}(b) is satisfied.
As a result, $A_{g}+iB_{3u}$ representation of superconductivity in UTe$_2$ ensures the presence of the quantum-geometry-induced anapole superconductivity regardless of the size of the interband pairing.

\begin{figure}[tbp]
    \centering
    \includegraphics[width=0.5\textwidth]{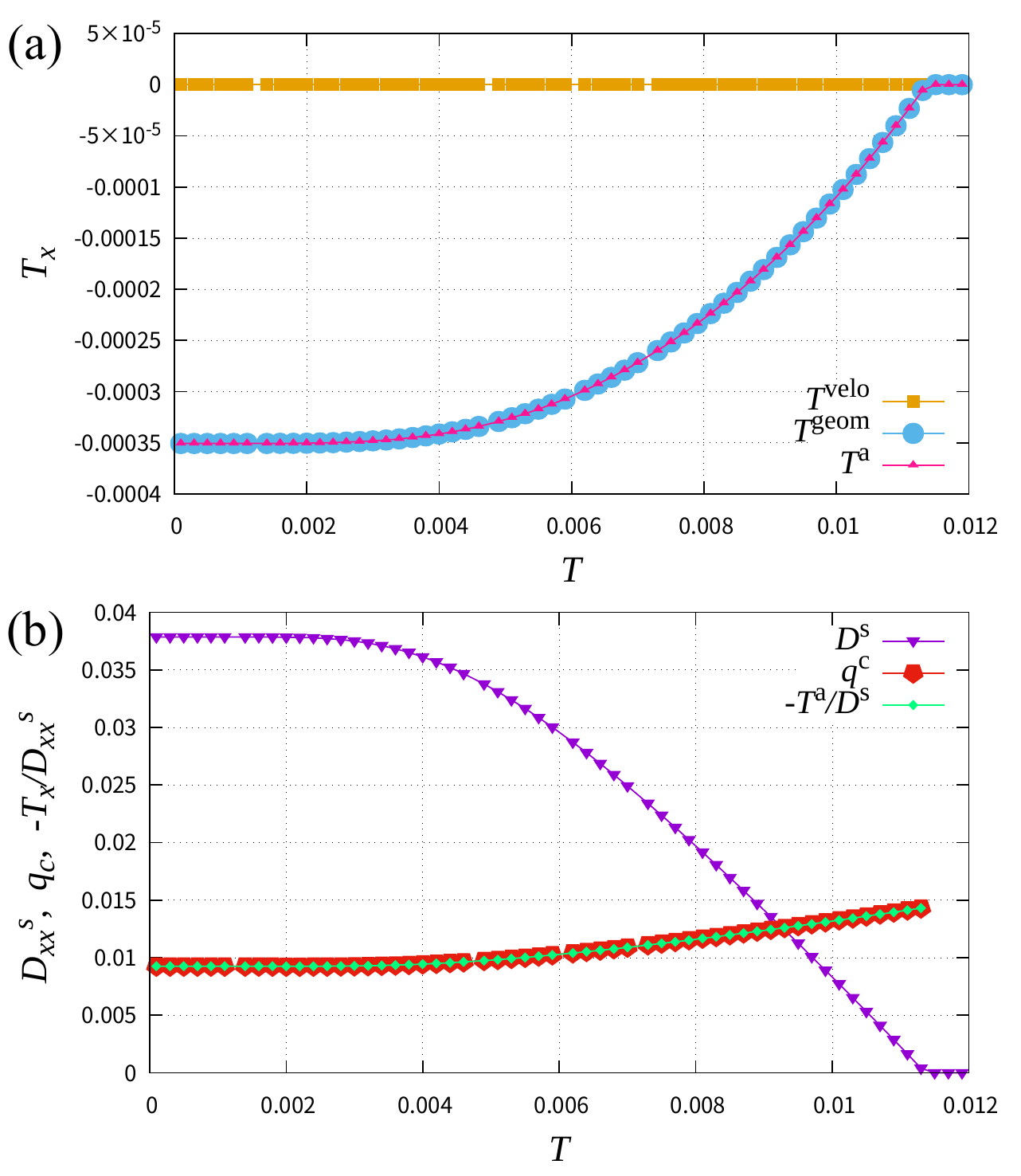}
    \caption{The temperature dependence of the anapole moment for the pair potential Eqs.~\eqref{eq:phig} and \eqref{eq:duz}.
    (a) The orange, blue, and pink lines show the group velocity term, geometric term, and total anapole moment, i.e. $T^{\rm velo}_x, T^{\rm geom}_x$, and $T_x$, respectively. The blue and pink lines coincide because $T^{\rm geom}_x =T_x$ in this case.
    (b) The purple and red lines show the superfluid weight $D^{\rm s}_{xx}$ and the most stable center of mass momenta of Cooper pairs $q_{\rm c}$.
    We also show $-T_x/D^{\rm s}_{xx}$ %\approx q_{\rm c}$ 
    by the green line, which almost coincides with the red line for $q_{\rm c}$.}
    \label{fig:ana_phig_duz_hole_ap}
\end{figure}

We calculate the anapole moment by Eqs.~\eqref{eq:anapole_two_term}-\eqref{eq:geom} and show the temperature dependence of $T_{x}$ in Fig.~\ref{fig:ana_phig_duz_hole_ap}(a).
Note that $T_{y} =  T_{z} = 0$ owing to the symmetry of $A_{{\rm g}}+iB_{3{\rm u}}$ irreducible representation (Table~\ref{table:symmetry_analysis}). The anapole moment does not depend on the sign of $\alpha$ in this case.
We denote the total anapole moment as $T^{\rm a}_{\mu}$ in all figures to avoid the confusion with the temperature $T$.
We find that the group velocity term (orange line) is always zero, revealing that the prediction based on the GL theory is exact. Therefore, the geometric term (blue line)
determines the anapole moment (pink line).
In Fig.~\ref{fig:ana_phig_duz_hole_ap}(b), we also plot the superfluid weight $D^{\rm s}_{xx}$, defined as the second-order $\bm q$-derivative of the free energy at $\bm q = 0$ (see Appendix~\ref{appendix:sfw_qc} for details). %(Fig.~\ref{fig:ana_phig_duz_hole_ap}(b)). 
We can evaluate the center of mass momentum of Cooper pairs in the anapole superconducting state by $-T_{x}/D^{\rm s}_{xx}$ (see Appendix~\ref{appendix:sfw_qc}). %which is shown by the green line in Fig.~\ref{fig:ana_phig_duz_hole_ap}(b). 
We also directly calculate the center of mass momentum $q_{\rm c}$, which minimizes the free energy, and compare it %(red line in Fig.~\ref{fig:ana_phig_duz_hole_ap}(b)) 
with $-T_{x}/D^{\rm s}_{xx}$ in Fig.~\ref{fig:ana_phig_duz_hole_ap}(b). %(green line).
%is also shown by the red line. 
%red line in Fig.~\ref{fig:ana_phig_duz_hole_ap}(b), can be estimated by $- T_{x}/D^{\rm s}_{xx}$ when $q_c$ is small (see Appendix~\ref{appendix:sfw_qc} for more details).
Since $-T_{x}/D^{\rm s}_{xx}$ almost coincides with $q_{\rm c}$, we confirm that $-T_{x}/D^{\rm s}_{xx}$ provides a good estimation for $q_{\rm c}$, indicating that higher-order $\bm q$-derivatives can be ignored.
%and $q_c$ becomes suddenly finite accompanied by the superconducting transition.
Thus, we conclude that the finite-$\bm q$ pairing state due to the anapole superconductivity is determined by the anapole moment. We stress that the asymmetric BS does not appear in this model, and the anapole superconductivity has a purely quantum geometric origin. More specifically, the Berry connection of Bloch electrons and the interband pairing play essential roles. 
%the anapole superconductivity can be easily induced by the Berry connection and the interband pairing in UTe$_2$. 

\begin{figure*}[htbp]
    \centering
    \includegraphics[width=1.0\textwidth]{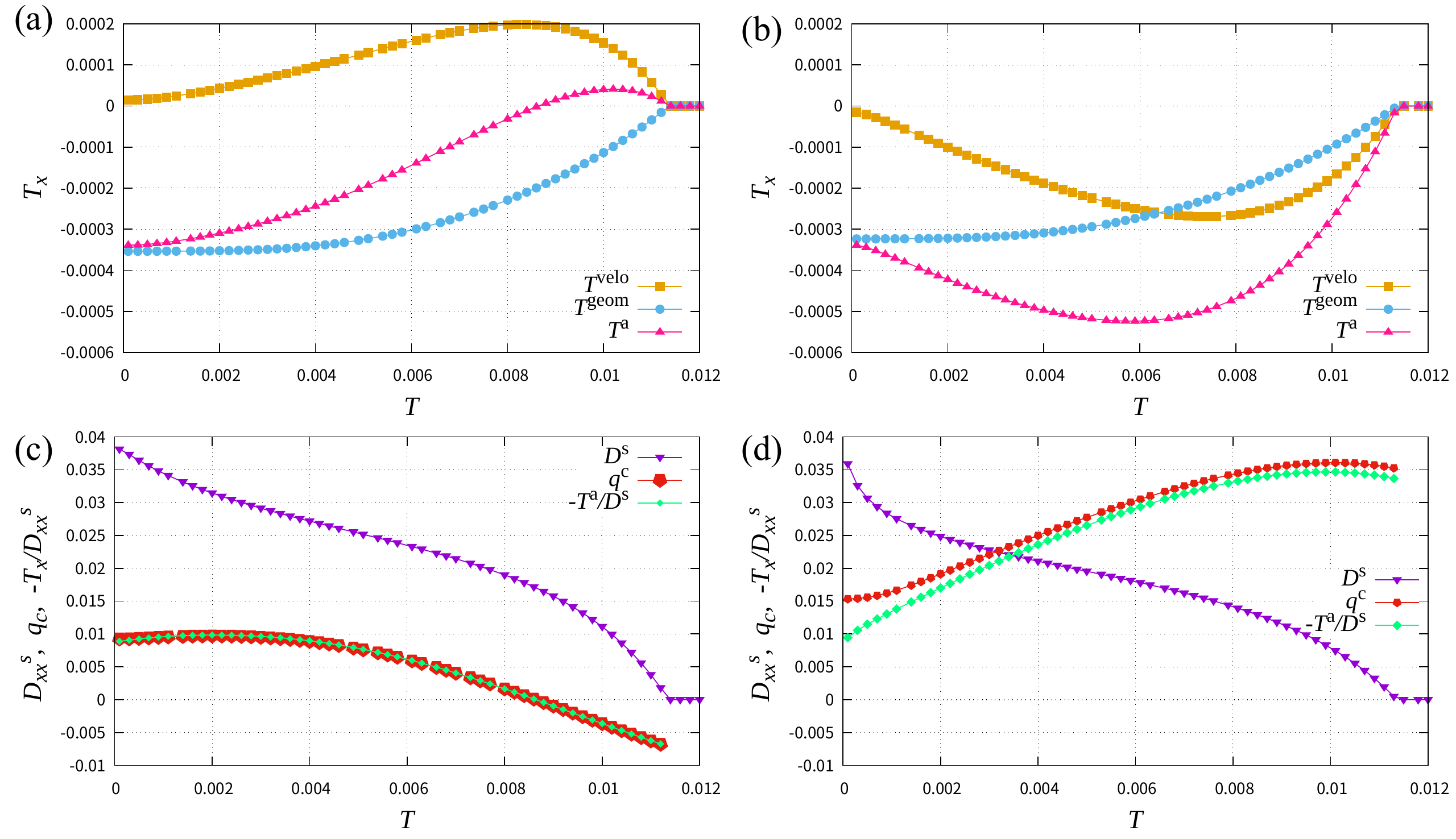}
    \caption{(a) (b) The anapole moment and (c) (d) the superfluid weight and $q_{\rm c}$ for the pair potential Eqs.~\eqref{eq:phig_dgy} and \eqref{eq:duz2}.
    We assume $\alpha=0.04$ in the panels (a) and (c), while $\alpha=-0.04$ in (b) and (d). 
    %The meaning of lines with colors is the same as
    The lines with colors show the same quantities as in
    Fig.~\ref{fig:ana_phig_duz_hole_ap}
    \label{fig:ana_phig_dgy_duz_hole}}
\end{figure*}

Next, we consider the superconducting pair potential,
\begin{align}
    &\bm \Delta^{\rm g}(T) = \Delta_{0}(T)\left(\phi_{\rm g}^{0}\sigma_0\otimes\tau_0+d_{{\rm g},y}^{z}\sigma_y\otimes\tau_z\right),\label{eq:phig_dgy}\\
    &\bm \Delta^{\rm u}(T) = i\Delta_{0}(T)d_{{\rm u},z}^{0}\sigma_z\otimes\tau_0,\label{eq:duz2}
\end{align}
with $d_{{\rm g},y}^{z}= \sin k_x$, which also belongs to the $A_{{\rm g}}+iB_{3{\rm u}}$ irreducible representation.
%We note that the gap function has a line-node-like gap minimum while system is gapped.
Thus, $T_{y} = T_{z} = 0$ is satisfied as in the previous case. We assume the same temperature dependence of $\Delta_0(T)$ as before. 
In this case, the group velocity term becomes finite as expected based on the GL theory since the condition, $\partial(h_0\pm\vert\bm h\vert)\bm m_1\cdot\hat{\bm h}=\partial\epsilon_{\pm}\left(\bm d_{\rm g}^z\times\bm d_{\rm u}^0\right)\cdot\hat{\bm g} \neq 0$, in Table~\ref{table:ana_UTe2_condition}(a) is satisfied.
Similarly, the geometric term is also finite because $\bm m_1\cdot\partial\hat{\bm h}=\left(\bm d_{\rm g}^z\times\bm d_{\rm u}^0\right)\cdot\partial\hat{\bm g}$ and $\bm m_2\cdot\partial\hat{\bm h}=\phi_{\rm g}^0\left(\hat{\bm g}\times\bm d_{\rm u}^0\right)\cdot\partial\hat{\bm g}$ are finite.
As the group velocity term is first order in $\alpha$ according to the GL theory, we expect that the sign of the ASOC coupling constant $\alpha$ is essential for the group velocity term~\cite{groupcomment}.
In addition, a part of the geometric term due to $\bm m_1\cdot\partial\hat{\bm h}$ %=\left(\bm d_{\rm g}^z\times\bm d_{\rm u}^0\right)\cdot\partial\hat{\bm g}$, 
is the first-order term, and the sign of $\alpha$ also affects the geometric term.

In Figs.~\ref{fig:ana_phig_dgy_duz_hole}(a) and \ref{fig:ana_phig_dgy_duz_hole}(b), we show the temperature dependence of the anapole moment for $\alpha=0.04$ and $\alpha=-0.04$, respectively. 
We also show the superfluid weight and $q_{\rm c}$ %for $\alpha=0.04$ and $\alpha=-0.04$ 
in Figs.~\ref{fig:ana_phig_dgy_duz_hole}(c) and \ref{fig:ana_phig_dgy_duz_hole}(d). %respectively.
We find that the sign of $\alpha$ drastically changes the group velocity term, whose sign is opposite between $\alpha = 0.04$ and $\alpha = -0.04$.
The magnitude is different between the two cases, %of $\alpha = 0.04$ and $\alpha = -0.04$, 
provably due to an effect beyond the GL theory.
On the other hand, the sign of the Rashba ASOC $\alpha$ only slightly changes the geometric term.
Note that other physical quantities also depend on the sign of $\alpha$ since the band representation of the gap function depends on $\alpha$, as $\phi_{\rm g}^0+d_{{\rm g},y}^{z}g_{y}/\vert\bm g\vert$.
Actually, we see that the superfluid weight depends on the sign of $\alpha$.

In both Figs.~\ref{fig:ana_phig_dgy_duz_hole}(a) and \ref{fig:ana_phig_dgy_duz_hole}(b), the group velocity term decays in the low temperature regime, which is attributed to the fact that the group velocity term is induced by the asymmetric BS.
The asymmetric BS, namely, $E_{a\bm k}\neq E_{a-\bm k}$, leads to non-equivalent distribution of Bogoliubov quasiparticles, $f(E_{a\bm k})\neq f(E_{a-\bm k})$.
This effect mainly induces the group velocity term.
However, the Fermi distribution function is reduced to the step function in the low temperature region, $f(E_{a\bm k}) \simeq \theta(-E_{a\bm k})$, which leads to $f(E_{a\bm k}) \simeq f(E_{a-\bm k})$ in the gapped system and suppresses the group velocity term (see also Appendix~\ref{appendix:velocity}). Therefore, the anapole moment is mainly determined by the geometric term in the low temperature region.
%\YYS{On the other hand, this implies that the asymmetric-band structure near the zero-energy level, such as the BFS at $\bm q = 0$, enhances the group velocity term, which is shown in Appendix~\ref{appendix:velocity}. Thus, the anapole superconductivity is mainly induced by quantum geometry in the low-temperature regions when the system at $\bm q = 0$ does not have large node.}
On the other hand, the relation $\theta(E_{a\bm k}) = \theta(E_{a-\bm k})$ does not hold when the BFS are present. Thus, the group velocity term can be sizable at $T=0$ when large BFS appear at $\bm q = 0$. This case is shown in Appendix~\ref{appendix:velocity}.

\begin{figure*}[htbp]
    \centering
    \includegraphics[width=1.0\textwidth]{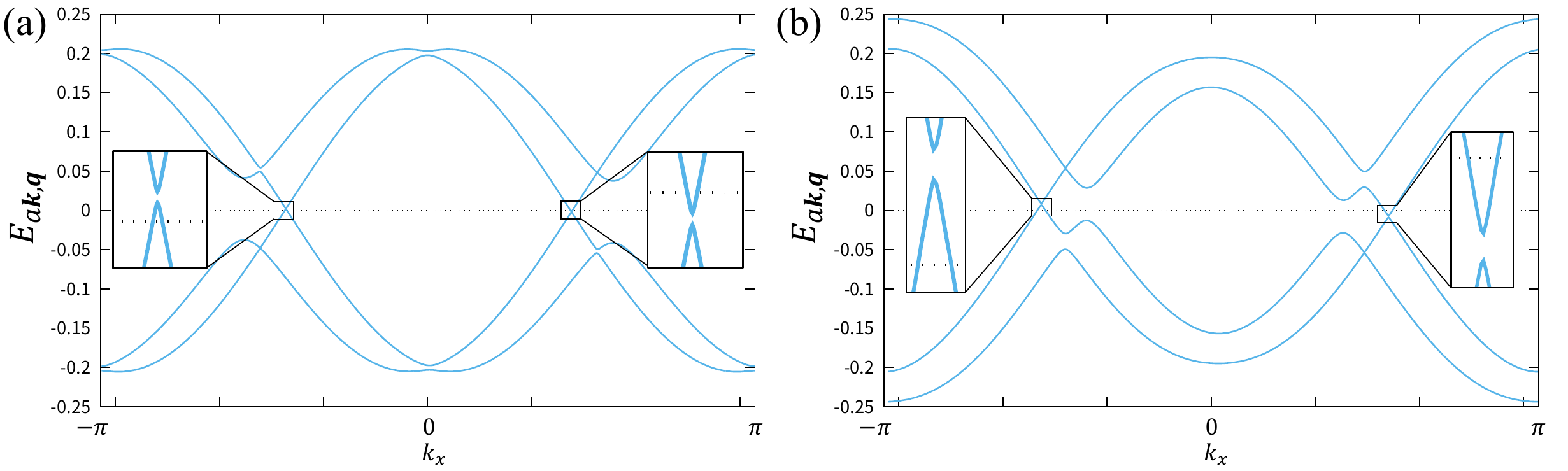}
    \caption{BS in the anapole superconducting state with ${\bm q} = q_{\rm c} \hat{x}$ %\TK{i.e. $E_{a\bm k,\bm q}$, }
    for the pair potential Eqs.~\eqref{eq:phig_dgy} and \eqref{eq:duz2}. Here, we plot $E_{a\bm k,\bm q}$ which follows the eigenvalue equation $H^{\rm BdG}_{\bm k,\bm q}\ket{\psi_{a\bm k,\bm q}}=E_{a\bm k,\bm q}\ket{\psi_{a\bm k,\bm q}}$. (a) BS on the line $(k_y,k_z)= (-0.0713998,0)$ for $\alpha=0.04$ and $T=0.002$. 
    (b) BS on the line $(k_y,k_z)= (0.499799,0)$ for $\alpha=-0.04$ and $T=0.01$. 
    The inset shows the enlarged view, which underlines the presence of BFS.
    }
    \label{fig:BS_phig_dgy_duz_hole_total}
\end{figure*}

In the model adopted in this section, while the BFS do not exist at $\bm q = 0$, they appear in the anapole superconducting state as a result of the center of mass momentum of Cooper pairs.
In Figs.~\ref{fig:BS_phig_dgy_duz_hole_total}(a) and \ref{fig:BS_phig_dgy_duz_hole_total}(b), we show the BS in the stable state ${\bm q} = q_{\rm c} \hat{x}$ for $\alpha = 0.04$ and $\alpha = -0.04$, respectively.
In the figures, the inset shows the presence of BFS.
In our model for $\bm q = 0$, the spin-triplet pairing component of the pair potential gives rise to the anisotropic gap structure. Therefore, %the system at  shows a tiny gap minimum (or node), and 
Bogoliubov quasiparticles with almost zero energy are present.
As a result, when the anapole moment tilts the BS along the direction of ${\bm q} = q_{\rm c} \hat{x}$, the BFS appear near the gap minimum.
Therefore, the anapole superconductivity can be verified by measuring the BFS.

\begin{figure}[tbp]
    \centering
    \includegraphics[width=0.5\textwidth]{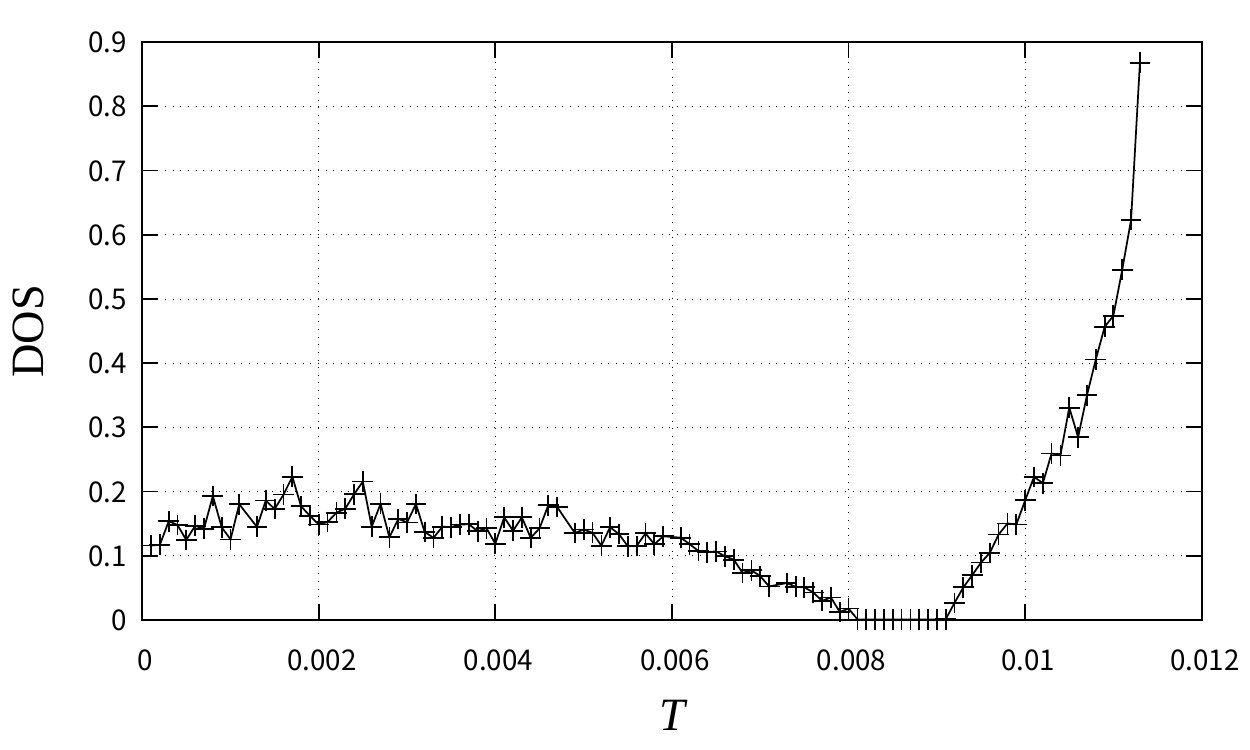}
    \caption{The temperature dependence of the DOS at the Fermi level calculated by $\sum_{\bm k}\sum_{a}\delta/\pi(\delta^2+E_{a\bm k, \bm q_{\rm c}}^2)$ with $\delta=1\times10^{-6}$.
    We assume the same parameters as Figs.~\ref{fig:ana_phig_dgy_duz_hole}(a) and \ref{fig:ana_phig_dgy_duz_hole}(c)}
%    We set $\alpha=0.04$ and assume the pair potential Eqs.~\eqref{eq:phig_dgy} and \eqref{eq:duz2}.}
    \label{fig:dos_phig_dgy_duz_hole_ap_total}
\end{figure}

Finally, we show an intriguing phenomenon originating from the competition of the group velocity and geometric terms in the anapole moment. 
In Figs.~\ref{fig:ana_phig_dgy_duz_hole}(a) and \ref{fig:ana_phig_dgy_duz_hole}(c), the anapole moment as well as $q_{\rm c}$ change the sign as the temperature decreases. This is because the geometric term has the opposite sign of the group velocity term and the group velocity term vanishes at the zero temperature.
As we mentioned above, the BFS are absent when the anapole moment is small $T_x \simeq 0$, while they appear for large $T_x$. Therefore, the BFS appear below $T=T_{\rm c}$, disappear in the intermediate temperature region, and reappear in the low temperature region by following the non-monotonic temperature dependence of the anapole moment. 
This is confirmed by the temperature dependence of DOS in Fig.~\ref{fig:dos_phig_dgy_duz_hole_ap_total}. %calculated by $\sum_{\bm k}\sum_{a}\delta^2/(\delta^2+E_{a\bm k}^2)$.
%Here, we set $\delta=1\times10^{-6}$.
The DOS at the Fermi level is zero around $T=0.0085$ since the anapole moment is small. On the other hand, the DOS is finite in the high and low temperature regions where the magnitude of the anapole moment is sizable.
This behavior is consistent with the reappearance of BFS, which is a characteristic feature of the anapole superconducting state with competing group velocity and geometric terms.
Thus, the role of quantum geometry on anapole superconductivity can be studied by measuring the zero energy DOS.

\section{Discussion\label{sec:discussion}}
In this paper, we showed some unique features of quantum-geometry-induced anapole superconductivity. A candidate material is UTe$_2$ and our result may pave the way for clarifying the symmetry of superconductivity in UTe$_2$. Thus, toward the experimental verification of anapole superconductivity in UTe$_2$, we give some discussions in this section.

A concern which is not limited to UTe$_2$ is the stability of finite-$\bm q$ pairing against the quantum fluctuation.
For the $s$-wave superconductivity in the isotropic and continuum model, the mean-field solution of the FFLO superconductivity is known to be unstable due to the quantum fluctuation~\cite{wang2018instability,zdybel2021stability}. This is attributed to the infinite degeneracy of finite-$\bm q$ pairing states ensured by the isotropic symmetry. In contrast, in the anapole superconductivity, the stable momentum of Cooper pairs $\bm q$ is restricted to only one direction, and additional degeneracy does not occur. Therefore, it is expected that anapole superconductivity is stable against quantum fluctuation. 
The argument is also supported by the fact that the finite-$\bm q$ pairing state is stable in the anisotropic three-dimensional system~\cite{zdybel2021stability}.
Our main target is the anisotropic three-dimensional systems, the case of UTe$_2$.

Then, we give some comments and remarks on the results relating to UTe$_2$, which include (1) the justification of the results obtained by the simplified model, (2) future issues, (3) methods for observing the anapole superconductivity, and (4) the relationship between our results and the recent experiments.

(1) In this paper, we dealt with a simplified model for UTe$_2$, where the orbital degree of freedom, electron correlation effect, detailed band structure, and so on are neglected for simplicity. 
Therefore, we discuss the situations in which our results are adaptable.
First, to derive the quantum-geometry-induced anapole superconductivity in UTe$_2$, we only assumed the presence of the ASOC due to the locally noncentrosymmetric structure as characteristic normal state property. %except for the presence of the polar-$s+ip$ state. 
Since the presence of the ASOC is universal for the crystal structure of UTe$_2$, we conclude that quantum-geometry-induced anapole superconductivity is also realized in more complicated models for UTe$_2$. Also, the decay of the group velocity term is universal when the BFS is absent for the zero center of mass momenta of Cooper pairs. Therefore, these results are adaptable for a wide range of the models for UTe$_2$.

Next, we comment on the BFS induced by the anapole moment and the sign change of the center of mass momenta, which are model-dependent. 
For the BFS to appear from the anapole moment, there must be gap minimum or node which depends on the Fermi surface and the order parameters. 
Such gap minimum or node is realistic and often obtained in microscopic calculations.
Especially, we would like to note that highly anisotropic momentum dependence in both even-parity and odd-parity pair potentials have been obtained based on the periodic Anderson model for UTe$_2$~\cite{ishizuka2021periodic}.
As for the sign change, competition between the geometric and group velocity terms is needed, which also depends on the band structure and order parameters.
In addition, in our model the presence or absence of the sign change depends also on the ASOC, which is hard to be predicted. 
As a result, the sign change of the center of mass momenta and the associated reentrant BFS depends on the model. %needs tuned parameter sets.
Thus, the BFS is likely to appear if the anapole superconductivity is realized in UTe$_2$, but the reappearing behavior of the BFS is not universal and should be verified by further calculations.

(2) From the above discussion, it is desired to study a more realistic model for UTe$_2$, as we have carried out for FeSe based on the first-principles calculation~\cite{kitamura2021thermodynamic,kitamura2021superconductivity}.
In a model taking account of various degrees of freedom, we may obtain significant contributions from the quantum geometry. Our previous study~\cite{kitamura2021superconductivity} has shown that the band degeneracy near the Fermi surfaces induces a large geometric contribution to the superfluid weight, implying that the band degeneracy may also be advantageous for the quantum-geometry-induced anapole superconductivity. 

It is also desired to solve multiple gap equations self-consistently, determining the amplitude and temperature dependence of two-component gap functions. While the necessary condition of the anapole superconductivity does not depend on such details, the self-consistent calculation of realistic models may enable quantitative estimation of the anapole moment, which, in turn, predicts the presence/absence of the sign change of the anapole moment.  
Such quantitative studies are beyond the scope of this paper and are left for future works.

(3) The anapole superconductivity can also be verified by other methods.
For example, the Josephson junction experiment, which was proposed to detect the helical superconductivity~\cite{kaur2005helical,josephcomment}, can apply to observe the anapole superconductivity. %which is suggested for observing 
In addition, a part of the authors proposed a unique vortex structure on anapole domains, current-induced anapole domain switching~\cite{kanasugi2022anapole}, and nonreciprocal optical and Meissner responses~\cite{watanabe2022optical,watanabe2022meissner}. 
%Recently, we also showed that the anapole moment is closely related to the superconducting piezoelectric effect~\cite{Chazono_SCPE,piezocomment}.
Recently, we also showed the superconducting piezoelectric effect~\cite{Chazono_SCPE} and intrinsic superconducting diode effect~\cite{daido2022intrinsic} in the anapole superconductors, which will be presented in another publication~\cite{piezocomment}. 

(4) We would like to stress that the symmetry of superconductivity in UTe$_2$ is unsettled. One of unresolved issues is the $\mathcal{T}$-symmetry breaking in the superconducting state, reported by the STM~\cite{jiao2020chiral} and the polar Kerr effect~\cite{hayes2021multicomponent}. Here we comment on the unidirectional property observed in the STM. It may be related to the anapole superconductivity, which is a unidirectional superconducting state in the bulk. Further studies are desired and ongoing to elucidate the exotic superconductivity in UTe$_2$. 

While we focused on the superconducting state at the zero magnetic field in this paper, an anapole superconducting state with finite-$\bm q$ pairing may also appear at finite magnetic fields.
The observation of the double superconducting transitions in UTe$_2$ under the magnetic field along the {\it b}-axis implies the superconducting phases with distinct symmetry~\cite{rosuel2022thermodynamic,sakai2022-field}.
%\TK{While $\mathcal{PT}$-symmetry is broken by the magnetic field in this phase, this symmetry breaking does not prohibit the $s+ip$-wave pairing which shares the symmetry with anapole or monopole state. Also in the case of anapole pairing under the magnetic field, finite-$\bm q$ pairing is allowed.}
If the superconducting state around $H_b \simeq 15$T is the $s+ip$-wave pairing state as discussed~\cite{rosuel2022thermodynamic}, it is either finite-$\bm q$ anapole superconductivity or monopole superconductivity. 
Although the $\mathcal{PT}$-symmetry is broken by the magnetic field in this phase, it does not suppress the finite-$\bm q$ pairing.
The possibility of anapole superconductivity under the magnetic field is also discussed in Ref.~\onlinecite{sakai2022-field}.

\section{Summary\label{sec:summary_discussion}}

In this paper, we showed that the quantum geometry of Bloch electrons induces the anapole superconductivity when the superconducting state breaks the $\mathcal{P}$- and $\mathcal{T}$-symmetry and has the polar symmetry. 
%the symmetry constraint is satisfied.
%The realization of the anapole superconductivity needs two inter-band effects in addition to the $\mathcal{P}$-, $\mathcal{T}$-symmetry breaking and polar pairing state.
Formulating the anapole moment characterizing the anapole superconductivity thoroughly, we find the group velocity term and geometric term with different origins. Based on the theory a model for UTe$_2$ was analyzed, and characteristic features of anapole superconductivity were clarified.

We identified microscopic processes for the group velocity term and geometric term of the anapole moment. At least two interband processes are necessary.
The previous study~\cite{kanasugi2022anapole} revealed that the asymmetric BS can induce the anapole superconductivity. This mechanism corresponds to the group velocity term. For the interband processes %need interband Cooper pairing. As a consequence, 
both even-parity and odd-parity pair potentials must have interband components. 
%corresponds to the case that two inter-band effects are given by the $\mathcal{PT}$-symmetric parity-mixed pair potential.
In contrast, the normal state Berry connection represents the interband process and gives rise to the geometric term.
Since quantum geometry arises from the geometric structure of Bloch wave functions, the geometric term does not need the asymmetric structure of BS.
Even when the odd structure of Cooper pairs does not affect the BS, the geometric term can be finite and cause the anapole superconductivity.
In other words, the quantum geometry can extract the odd structure of Cooper pairs which does not appear in the BS, and reflect it in the anapole moment.
Therefore, the anapole superconductivity may have a quantum geometric origin.
This case requires only either even-parity or odd-parity interband pair potential.
%in this case, anapole superconductivity only needs the presence of the Berry connection and one interband pairing, which is considered to be realized in UTe$_2$.

Furthermore, we clarified the general and unique features of anapole superconductivity.
First, in the low-temperature region, the anapole superconductivity is purely induced by the quantum geometry. This is because the group velocity term is suppressed by the superconducting gap. In other words, the quantum geometry is needed for the anapole superconductivity in the ground state.
Second, the BFS appears in the anapole superconducting state, when the gap is sufficiently anisotropic.
Third, when the group velocity term and the geometric term are competing, the anapole moment changes the sign as decreasing the temperature.
%This sign change is due to the competition geometric term and the group velocity term are different, 
This sign change
may lead to the nonmonotonic evolution of the BFS.
%While our model has the line-node-like gap minimum, this BFS can be also realized when system has point node or other type of gap minimum.
A candidate superconductor UTe$_2$ may host an anisotropic superconducting gap~\cite{Aoki_review2022}, and therefore, the appearance of the BFS is expected. Observation of the BFS and their unique temperature dependence would not only evidence the anapole superconductivity in UTe$_2$ but also provide a strong constraint on the symmetry of superconductivity (see Table~\ref{table:symmetry_analysis} for symmetry classification of parity-mixed superconducting states).
Therefore, search for BFS in UTe$_2$ is desirable.

In conclusion, the quantum geometry ubiquitously induces the anapole superconductivity in the $\mathcal{PT}$-symmetric parity-mixed pairing state in multiband superconductors. The anapole superconducting state may show unique phenomena which can be experimentally tested.
Thus, we propose a way to clarify the superconducting state in UTe$_2$.

\begin{acknowledgments}
We are grateful to A.~Daido, R.~Sano, D.~Aoki, J.-P.~Brison, K.~Ishida, Y.~Tokiwa, Y.~Tokunaga, and D.~F.~Agterberg for fruitful discussions.
This work was supported by JSPS KAKENHI (Grants Nos. JP18H01178, JP18H05227, JP20H05159, JP21K18145, JP22H01181, JP22J22520, JP22H04933), JST SPRING (Grant Number JPMJSP2110) and SPIRITS 2020 of Kyoto University.
\end{acknowledgments}

\appendix
\section{Derivation of anapole moment\label{sec:ana_derivation}}
In this section, we derive the anapole moment in the superconducting state.
We start from the BdG Hamiltonian written as,
\begin{eqnarray}
	\hat{H}^{\BdG} &=& \frac{1}{2}\sum_{\bm k}\hat{\tilde{\Psi}}^\dagger_{\bm k,\bm q}\tilde{H}^\BdG_{\bm k,\bm q}\hat{\tilde{\Psi}}_{\bm k,\bm q},\\
	\tilde{H}^\BdG_{\bm k,\bm q} &=& \left(
		\begin{array}{cc}
			H_{\bm k+\bm q}&\bm \Delta_{\bm k}U_{\mathcal{T}}%\left[i\sigma_y\otimes\bm 1\right]
			\\
			%\left[-i\sigma_y\otimes\bm 1\right]
			U_{\mathcal{T}}^\dagger\bm \Delta^\dagger_{\bm k}&-H_{-\bm k+\bm q}^T
		\end{array}
	\right), \\
	\hat{\tilde{\Psi}}^\dagger_{\bm k,\bm q} &=& \left(
	\begin{array}{cc}
		\hat{\bm c}^\dagger_{\bm k+\bm q}&\hat{\bm c}^T_{-\bm k+\bm q}
	\end{array}
\right).
\end{eqnarray}
Using the unitary operator, 
\begin{eqnarray}
   U_{\mathcal{T}_{\rm hole}} &=& \left(
		\begin{array}{cc}
			\sigma_0\otimes\bm 1&0\\
			0&U_{\mathcal{T}}
		\end{array}
	\right),
\end{eqnarray}
we can rewrite the BdG Hamiltonian as,
\begin{eqnarray}
    \hat{H}^{\BdG} &=& \frac{1}{2}\sum_{\bm k}\hat{\tilde{\Psi}}^\dagger_{\bm k,\bm q}U^\dagger_{\mathcal{T}_{\rm hole}}U_{\mathcal{T}_{\rm hole}}\tilde{H}^\BdG_{\bm k,\bm q}U^\dagger_{\mathcal{T}_{\rm hole}}U_{\mathcal{T}_{\rm hole}}\hat{\tilde{\Psi}}_{\bm k,\bm q},\notag\\
    &=& \frac{1}{2}\sum_{\bm k}\hat{\Psi}^\dagger_{\bm k,\bm q} H^\BdG_{\bm k,\bm q}\hat{\Psi}_{\bm k,\bm q}.
\end{eqnarray}
Thus, we define the Nambu Green function with Matsubara frequency $\omega_n$ as, $\mathcal{G}^\BdG_{\bm k,\bm q,\omega_n} =\left[i\omega_n-H^\BdG_{\bm k,\bm q}\right]^{-1}$.
Using this, the free energy is obtained as $F_{\bm q} = -\frac{1}{2\beta}\sum_{\bm k\omega_n}{\rm Tr}\ln[\mathcal{G}_{\bm k,\bm q,\omega_n}^{\BdG-1}]$, where ${\rm Tr}$ represents the trace over all internal degrees of freedom.
The anapole moment is defined as the first-order coefficient of the superconducting free energy with respect to $\bm q$~\cite{kanasugi2022anapole}:
\begin{eqnarray}
    T_{\mu} &=& \lim_{\bm q\rightarrow0}\dfrac{dF_{\bm q}}{dq_\mu},\notag\\
    &=& \lim_{\bm q\rightarrow0}\left(\partial_{q_\mu}F_{\bm q}+\sum_l\partial_{q_\mu}\Delta_l(\bm q)\partial_{\Delta_l(\bm q)}F(\bm q)\right),\notag\\
    &=& \lim_{\bm q\rightarrow0}\partial_{q_\mu}F_{\bm q},\notag\\
    &=& \dfrac{1}{2\beta}\lim_{\bm q\rightarrow0}\sum_{\bm k\omega_n}{\rm Tr}\left[\mathcal{G}^\BdG_{ \bm k,\bm q,\omega_n}\partial_{q_\mu}H^\BdG_{\bm k,\bm q}\right],\notag\\
    &=& \dfrac{1}{2\beta}\sum_{\bm k\omega_n}{\rm Tr}\left[\mathcal{G}^\BdG_{ \bm k,\omega_n}\partial_{\mu}H^+_{\bm k}\right],
\end{eqnarray}
where $\mathcal{G}^\BdG_{ \bm k,\omega_n} = \mathcal{G}^\BdG_{ \bm k,\bm q,\omega_n}\vert_{\bm q = 0}$.
Here, we use the relationship $\partial_{\Delta_l(\bm q)}F(\bm q) = 0$, in which $\Delta_l(\bm q)$ denotes the $\bm k$-independent part of each component of the gap function, since the superconducting state is stable.
Taking the sum of the Matsubara frequencies, we get the formula of the anapole moment as,
\begin{eqnarray}
    	T_{\mu} &=& \dfrac{1}{2}\sum_{\bm k}\sum_af(E_{a\bm k})
	\bra{\psi_{a\bm k}}\partial_\mu H^+_{\bm k}\ket{\psi_{a\bm k}}.
\end{eqnarray}

\section{Anapole moment from GL theory\label{appendix:ana_gl}}
%For the GL expansion, first, we rewrite the free energy as,
%\begin{eqnarray}
%    F_{\bm q} &=& -\frac{1}{2}T\sum_{\bm k\omega_n}{\rm Tr}\ln[\mathcal{G}_{\BdG\bm k,\bm q,\omega_n}^{-1}]\notag\\
%    &=& -\frac{1}{2}T\sum_{\bm k\omega_n}{\rm Tr}\ln[\mathcal{G}_{\bm k,\bm q,\omega_n}^{-1}-(1-\mathcal{G}_{\bm k,\bm q,\omega_n}\mathcal{V}_{\bm k})],\label{eq:free_ex_gl}
%\end{eqnarray}
%where,
%\begin{eqnarray}
%    &\mathcal{V}_{\bm k} = \left(
%        \begin{array}{cc}
%            0 & \bm \Delta_{\bm k} \\
%            \bm \Delta^\dagger_{\bm k} & 0
%        \end{array}
%    \right),\\
%    &\mathcal{G}_{\bm k,\bm q,\omega_n}^{-1}=\mathcal{G}_{\BdG\bm k,\bm q,\omega_n}^{-1}+\mathcal{V}_{\bm k}.
%\end{eqnarray}
Here, we derive the formula of the anapole moment using the GL expansion.
Since we focus on the superconducting state, we ignore the free-electron term and rewrite the free energy as, 
\begin{eqnarray}
    F_{\bm q} &=&  -\frac{1}{2\beta}\sum_{\bm k\omega_n}{\rm Tr}\ln[\mathcal{G}_{\bm k,\bm q,\omega_n}^{\BdG-1}],\notag\\
    &=&  -\frac{1}{2\beta}\sum_{\bm k\omega_n}{\rm Tr}\left[\ln[1-\mathcal{G}_{\bm k,\bm q,\omega_n}\mathcal{V}_{\bm k}]+\ln[\mathcal{G}_{\bm k,\bm q,\omega_n}^{-1}]\right],\notag\\
    &=&  \frac{1}{2\beta}\sum_{c=1}^{\infty}\sum_{\bm k\omega_n}{\rm Tr}[\mathcal{G}_{\bm k,\bm q,\omega_n}\mathcal{V}_{\bm k}\mathcal{G}_{\bm k,\bm q,\omega_n}\mathcal{V}_{\bm k}]^c+\cdots,
    %\frac{1}{2\beta}\sum_{c}\sum_{\bm k\omega_n}{\rm Tr}[\mathcal{G}_{\bm k,\bm q,\omega_n}\mathcal{V}_{\bm k}\mathcal{G}_{\bm k,\bm q,\omega_n}\mathcal{V}_{\bm k}]^c+\cdots
\end{eqnarray}
where,
\begin{eqnarray}
    &\mathcal{V}_{\bm k} = \left(
        \begin{array}{cc}
            0 & \bm \Delta_{\bm k} \\
            \bm \Delta^\dagger_{\bm k} & 0
        \end{array}
    \right),\\
    &\mathcal{G}_{\bm k,\bm q,\omega_n}^{-1}=\mathcal{G}_{\bm k,\bm q,\omega_n}^{\BdG-1}+\mathcal{V}_{\bm k}.
\end{eqnarray}
Up to the second order of the superconducting order parameter, the free energy can be written as,
\begin{eqnarray}
    F^{\rm GL}_{\bm q} = \frac{1}{2\beta}\sum_{\bm k\omega_n}&{\rm Tr}&\left[\mathcal{G}_{\bm k,\bm q,\omega_n}\mathcal{V}_{\bm k}\mathcal{G}_{\bm k,\bm q,\omega_n}\mathcal{V}_{\bm k}\right],\notag\\
    %\frac{1}{2\beta}\sum_{c}\sum_{\bm k\omega_n}{\rm Tr}[\mathcal{G}_{\bm k,\bm q,\omega_n}\mathcal{V}_{\bm k}\mathcal{G}_{\bm k,\bm q,\omega_n}\mathcal{V}_{\bm k}]\notag\\
    = \dfrac{1}{2\beta}\sum_{\bm k\omega_n}&{\rm tr}&[\mathcal{G}^{\rm p}_{\bm k,\bm q,\omega_n}\bm \Delta_{\bm k}\mathcal{G}^{\rm h}_{\bm k,\bm q,\omega_n}\bm \Delta^\dagger_{\bm k}\notag\\
    &+&\mathcal{G}^{\rm h}_{\bm k,\bm q,\omega_n}\bm \Delta^\dagger_{\bm k}\mathcal{G}^{\rm p}_{\bm k,\bm q,\omega_n}\bm \Delta_{\bm k}],\notag\\
    = \dfrac{1}{\beta}\sum_{\bm k\omega_n}&{\rm tr}&\left[\mathcal{G}^{\rm p}_{\bm k,\bm q,\omega_n}\bm \Delta_{\bm k}\mathcal{G}^{\rm h}_{\bm k,\bm q,\omega_n}\bm \Delta^\dagger_{\bm k}\right],\label{eq:free_gl}
\end{eqnarray}
with $\mathcal{G}_{\bm k,\bm q,\omega_n}^{{\rm p(h)}-1} = i\omega_n\mp H_{\bm k\pm\bm q}$.
Then, expanding Eq.~\eqref{eq:free_gl} with respect to the $\bm q$ up to the first order, we get,
\begin{align}
    F^{\rm GL}_{\bm q} 
    = \dfrac{1}{\beta}&\sum_{\bm k\omega_n}{\rm tr}[\partial_\mu\mathcal{G}^{\rm p}_{\bm k,\omega_n}\bm \Delta_{\bm k}\mathcal{G}^{\rm h}_{\bm k,\omega_n}\bm \Delta_{\bm k}^\dagger\notag\\
    &-\mathcal{G}^{\rm p}_{\bm k,\omega_n}\bm \Delta_{\bm k}\partial_\mu\mathcal{G}^{\rm h}_{\bm k,\omega_n}\bm \Delta_{\bm k}^\dagger]q_\mu+\cdots,\notag\\
    = \dfrac{1}{\beta}&\sum_{\bm k\omega_n}{\rm tr}[\mathcal{G}^{\rm p}_{\bm k,\omega_n}\partial_\mu H_{\bm k}\mathcal{G}^{\rm p}_{\bm k,\omega_n}\bm \Delta_{\bm k}\mathcal{G}^{\rm h}_{\bm k,\omega_n}\bm \Delta_{\bm k}^\dagger\notag\\
    &+\mathcal{G}^{\rm p}_{\bm k,\omega_n}\bm \Delta_{\bm k}\mathcal{G}^{\rm h}_{\bm k,\omega_n}\partial_\mu H_{\bm k}\mathcal{G}^{\rm h}_{\bm k,\omega_n}\bm \Delta_{\bm k}^\dagger]q_\mu+\cdots.
\end{align}
Here, we use $\partial_{\mu} \mathcal{G}^{\rm p(h)}_{\bm k,\omega_n} = (-)\mathcal{G}^{\rm p(h)}_{\bm k,\omega_n}\partial_{\mu} H_{\bm k} \mathcal{G}^{\rm p(h)}_{\bm k,\omega_n}$.
%Thus, the coefficient of $q_\mu$ is the effective anapole moment in the GL theory.
%Using the relationship, $G_{p\bm k\omega_n} = -G_{p\bm k-\omega_n}$, we can get ,
Using the relationship, $\mathcal{G}^{\rm p}_{\bm k\omega_n} = -\mathcal{G}^{\rm h}_{\bm k-\omega_n}$, we obtain
\begin{align}
    T^{\rm GL}_{\mu} = \dfrac{1}{\beta}\sum_{\bm k\omega_n}&{\rm tr}
	\left[\mathcal{G}^{\rm p}_{\bm k\omega_n}\partial_\mu H_{\bm k}\mathcal{G}^{\rm p}_{\bm k\omega_n}\bm \Delta_{\bm k}\mathcal{G}^{\rm h}_{\bm k\omega_n} \bm \Delta^\dagger_{\bm k}\right.\notag\\
	&-\left.
	\mathcal{G}^{\rm p}_{\bm k\omega_n}\partial_\mu H_{\bm k}\mathcal{G}^{\rm p}_{\bm k\omega_n} \bm \Delta^\dagger_{\bm k}\mathcal{G}^{\rm h}_{\bm k\omega_n}\bm \Delta_{\bm k}
	\right]~\label{eq:gl_ana_all}.
\end{align}

In the remaining part of this section, we discuss the symmetry constraint on the anapole moment and simplify Eq.~\eqref{eq:gl_ana_all}.
We assume that the normal state Hamiltonian is $\mathcal{P}$- and $\mathcal{T}$-symmetric, i.e, $H_{\bm k}\xrightarrow{\mathcal{P}}U_{\mathcal{P}} H_{-\bm k} U_{\mathcal{P}}^\dagger=H_{\bm k}$ and $H_{\bm k}\xrightarrow{\mathcal{T}}U_{\mathcal{T}} H^*_{-\bm k} U_{\mathcal{T}}^\dagger=H_{\bm k}$, where $U_{\mathcal{P}}$ is the unitary operator for the $\mathcal{P}$-symmetry.
Here, we require that the $\mathcal{T}$ operator commutes with the $\mathcal{P}$ operator, i.e. $U_{\mathcal{T}}\mathcal{K}U_{\mathcal{P}}=U_{\mathcal{T}}U_{\mathcal{P}}^*\mathcal{K}=U_{\mathcal{P}}U_{\mathcal{T}}\mathcal{K}$ with complex conjugate operator $\mathcal{K}$ and the $\mathcal{P}$ operator is its own inverse, i.e. $U_{\mathcal{P}}^2 =\bm 1$.
In addition, the $\mathcal{T}$ operator satisfies $U_{\mathcal{T}}\mathcal{K}U_{\mathcal{T}}\mathcal{K}=U_{\mathcal{T}}U^*_{\mathcal{T}}=-\bm 1$, since we consider spinful electron systems.

Let us consider the $\mathcal{T}$-symmetry in the superconducting state. 
Under the $\mathcal{T}$ operation, the pair potential follows $\bm \Delta_{\bm k}U_{\mathcal{T}}\xrightarrow{\mathcal{T}} U_{\mathcal{T}}\bm \Delta^*_{-\bm k}U^*_{\mathcal{T}}U^T_{\mathcal{T}}$.
The fermionic anti-symmetry, $\bm \Delta_{\bm k}U_{\mathcal{T}}=-U_{\mathcal{T}}^T\bm \Delta^T_{-\bm k}$, leads to $U_{\mathcal{T}}\bm \Delta^*_{-\bm k}U^*_{\mathcal{T}}U^T_{\mathcal{T}}=-\bm \Delta^\dagger_{\bm k}U^T_{\mathcal{T}}=\bm \Delta^\dagger_{\bm k} U_{\mathcal{T}}$ since $U_{\mathcal{T}}U^*_{\mathcal{T}}=-\bm 1$ is satisfied, which means $\bm \Delta_{\bm k}\xrightarrow{\mathcal{T}}\bm \Delta_{\bm k}^\dagger$.
As a result, when the pair potential is $\mathcal{T}$-symmetric, the first term of Eq.~\eqref{eq:gl_ana_all} cancels out the second term, and therefore, the anapole moment vanishes.

Next, we consider the $\mathcal{P}$-symmetry. 
%For the space inversion, ]
Since $U_{\mathcal{T}}U_{\mathcal{P}}^*=U_{\mathcal{P}}U_{\mathcal{T}}$ and $U_{\mathcal{P}}^2 =\bm 1$ lead to $U_{\mathcal{P}}U_{\mathcal{T}}=U_{\mathcal{T}}U_{\mathcal{P}}^T$, the pair potential follows $\Delta_{\bm k}U_{\mathcal{T}}\xrightarrow{\mathcal{P}} U_{\mathcal{P}}\bm \Delta_{-\bm k}U_{\mathcal{T}}U^T_{\mathcal{P}}=U_{\mathcal{P}}\bm \Delta_{-\bm k}U^\dagger_{\mathcal{P}}U_{\mathcal{T}}$, which means $\bm \Delta_{\bm k}\xrightarrow{\mathcal{P}}U^\dagger_{\mathcal{P}}\bm \Delta_{-\bm k}U_{\mathcal{P}}=\bm \Delta_{{\rm g}\bm k}-\bm \Delta_{{\rm u}\bm k}$.
Therefore, because of $U_{\mathcal{P}}\mathcal{G}^{\rm p(h)}_{-\bm k\omega_n}U^\dagger_{\mathcal{P}}=\mathcal{G}^{\rm p(h)}_{\bm k\omega_n}$ and $U_{\mathcal{P}}\partial_{-\mu} H_{-\bm k}U^\dagger_{\mathcal{P}}=-\partial_{\mu} H_{\bm k}$ with $\partial_{-\mu} = \frac{\partial}{\partial(-k_\mu)}$, the $\mathcal{P}$-even part of the effective anapole moment,
\begin{align}
    T^{\rm GL:even}_{\mu} &= \dfrac{1}{\beta}\sum_{\bm k\omega_n}{\rm tr}
	\left[\mathcal{G}^{\rm p}_{\bm k\omega_n}\partial_\mu H_{\bm k}\mathcal{G}^{\rm p}_{\bm k\omega_n}\bm \Delta^{\rm g}_{\bm k}\mathcal{G}^{\rm h}_{\bm k\omega_n} \bm \Delta^{{\rm g}\dagger}_{\bm k}\right.\notag\\
	&-\left.
	\mathcal{G}^{\rm p}_{\bm k\omega_n}\partial_\mu H_{\bm k}\mathcal{G}^{\rm p}_{\bm k\omega_n} \bm \Delta^{{\rm u}\dagger}_{\bm k}\mathcal{G}^{\rm h}_{\bm k\omega_n}\bm \Delta^{\rm u}_{\bm k}
	\right]+({\rm g}\leftrightarrow{\rm u}) %~\label{eq:gl_ana_all},
\end{align}
vanishes due to the cancellation between $\bm k$ and $-\bm k$.
Thus, only the $\mathcal{P}$-odd and $\mathcal{T}$-odd part of the anapole moment becomes finite and we arrive at  Eq.~\eqref{eq:ana_gl} in the main text.

\section{Interband effect on anapole moment \label{appendix:degenerate}}

Here, we show that at least two interband processes are needed for the anapole moment.
We start from Eq.~\eqref{eq:ana_gl_b} based on the GL theory and consider the contribution from the purely intraband process, namely, the case $n = m = p$.
Below, the $\bm k$ dependence is omitted for simplicity.
When $\chi_n =\chi_m=\chi_p$ is satisfied, the first and second terms of Eq.~\eqref{eq:ana_gl_b} obviously cancel out each other.
Therefore, we consider the other cases.

First, when we fix the U(1) gauge of the Bloch wave function, we can define the relationship between the Kramers doublet through the $\mathcal{PT}$ symmetry as
$    
U_\mathcal{PT}\ket{u_{n\uparrow}^*} = \ket{u_{n\downarrow}}
$ with $U_{\mathcal{PT}}=U_\mathcal{P}U_\mathcal{T}$;
this leads to
$
-\ket{u_{n\uparrow}} = U_\mathcal{PT}\ket{u_{n\downarrow}^*}
$ because of  $U_\mathcal{PT}U_\mathcal{PT}^* = -\bm 1$.
As a result, in the case of $\chi_n \neq \chi_m $, the velocity operator of the normal state vanishes as follows:
\begin{eqnarray}
    \bra{u_{n\chi_n}}\partial H\ket{u_{n\chi_m}}
    &=&\left(\bra{u_{n\chi_m}}\partial H\ket{u_{n\chi_n}}\right)^*\notag\\
    &=&\bra{u_{n\chi_m}^*}\partial H^*\ket{u_{n\chi_n}^*}\notag\\
    &=&\bra{u_{n\chi_m}^*} U^\dagger_{\mathcal{PT}}U_{\mathcal{PT}}\partial H^*U^\dagger_{\mathcal{PT}}U_{\mathcal{PT}}\ket{u_{n\chi_n}^*}\notag\\
    &=& -\bra{u_{n\chi_n}}\partial H\ket{u_{n\chi_m}}\notag\\
    &=& 0,
\end{eqnarray}
since either $\chi_n$ or $\chi_m$ corresponds to the state $\downarrow$.
Therefore, the contribution to Eq.~\eqref{eq:ana_gl_b} vanishes, and we have only to consider the rest case, $\chi_n = \chi_m \neq \chi_p$.
In this case, contribution to Eq.~\eqref{eq:ana_gl_b} from each $\k$ and $\omega_n$ can be written as, 
\begin{align}
    &\dfrac{1}{\beta}\sum_{n,\chi_n\neq\chi_p}C^{\rm GL}_{nnn}
	\partial \epsilon_n
	\left(
	\bra{u_{n\chi_n}}\bm \Delta^{\rm g}
	\ket{u_{n\chi_p}}
	\bra{u_{n\chi_p}}
	\bm \Delta^{{\rm u}\dagger}\ket{u_{n\chi_n}}
	\right.\notag\\
	&\left.
	+\bra{u_{n\chi_p}}\bm \Delta^{{\rm g}\dagger}_{\bm k}\ket{u_{n\chi_n}}
	\bra{u_{n\chi_n}}\bm \Delta^{\rm u}_{\bm k}\ket{u_{n\chi_p}}
	\right)-({\rm g}\leftrightarrow{\rm u}).\label{eq:dege_velo}
\end{align}
Because of $\bm \Delta^{\rm g} = \bm \Delta^{\rm g\dagger}$ and $\bm \Delta^{\rm u} = -\bm \Delta^{\rm u\dagger}$ except for the U(1)-gauge dependence of Cooper pairs, the first and second terms of Eq.~\eqref{eq:dege_velo} cancel out each other.
Thus, we found that the purely intraband process can not produce the anapole moment. This means that at least two interband effects are needed for the anapole superconductivity, as we discuss in Sec.~\ref{sec:origin_general}.

\section{Group velocity term, asymmetric BS, and BFS\label{appendix:velocity}}

In this section, we discuss the group velocity term of anapole moment.
The following discussions are based on the general two-band model introduced in Sec.~\ref{sec:two-band_model}.

First, we show the close relationship between the group velocity term and asymmetric BS and elucidate the mechanism of the decay of the group velocity term in the low temperature region.
Assuming $h_{0\bm k}\gg\vert \bm h_{\bm k}\vert$, we approximate the velocity operator in the BdG form,
\begin{eqnarray}
    \partial_{\mu} H^+_{\bm k} = \partial_\mu h_{0\bm k}\bm 1.
\end{eqnarray}
When the polar direction is denoted as $\mu$, the anapole moment is obtained from Eq.~\eqref{eq:anapole_moment} as,
\begin{eqnarray}
    T_{\mu} &=& \dfrac{1}{2}\sum_{\bm k}\sum_af(E_{a\bm k})
	\partial_\mu h_{0\bm k},\notag\\
	&=& \dfrac{1}{2}\sum_{\bm k(k_\mu > 0)}\sum_a\left(f(E_{a\bm k})-f(E_{a-\bm k})\right)
	\partial_\mu h_{0\bm k},
\end{eqnarray}
which corresponds to the group velocity term since $\partial_{\mu}h_{0\bm k}$ is the group velocity.
Thus, $f(E_{a\bm k})-f(E_{a-\bm k})$, which represents the asymmetric structure of BS, induces the group velocity term.
However, the Fermi distribution function is approximated by the step function, $f(E_{a\bm k})\approx \theta(-E_{a\bm k})$, in the low temperature region, and therefore, $f(E_{a\bm k})-f(E_{a-\bm k}) = 0$ at the zero temperature when $E_{a\bm k}$ and $E_{a-\bm k}$ have the same sign.
Thus, the group velocity term vanishes at $T=0$, unless the BFS are present.

\begin{figure}[htbp]
    \centering
    \includegraphics[width=0.5\textwidth]{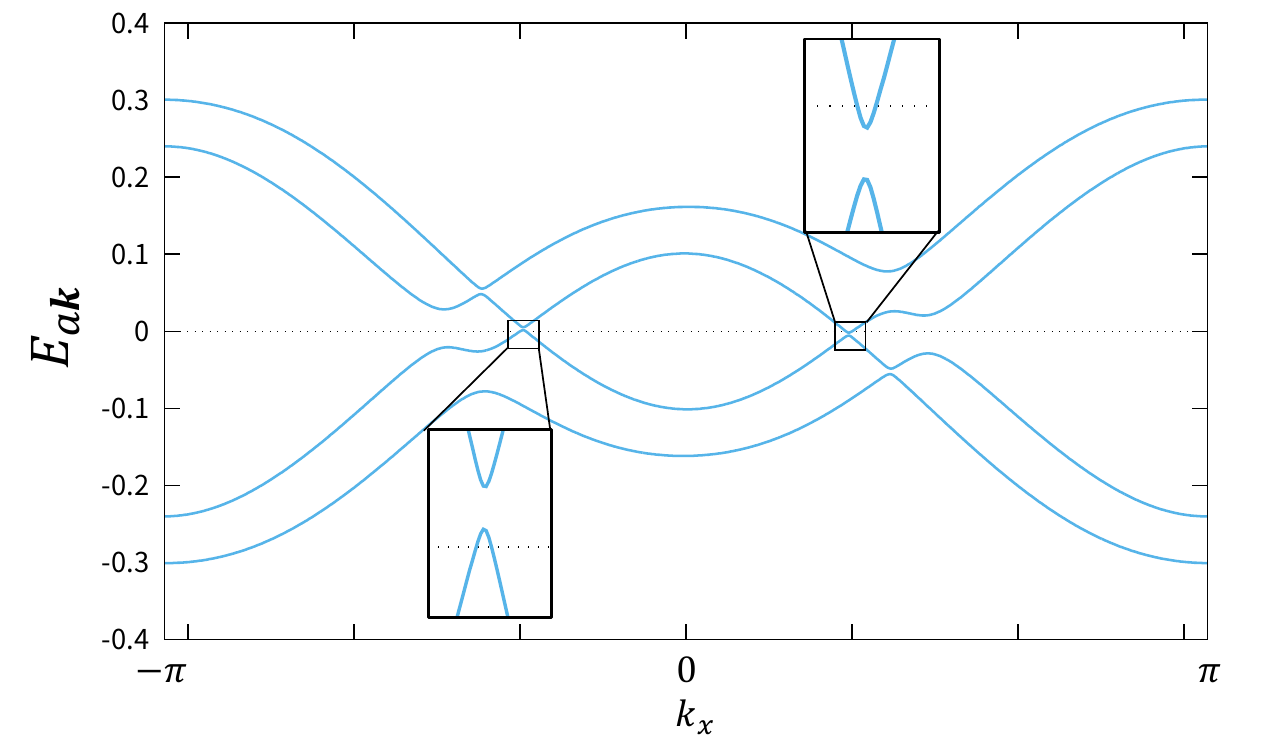}
    \caption{BS on the line $(k_y,k_z)= (-0.862398,0)$. We assume $\alpha=0.04$ and the pair potential in Eqs.~\eqref{eq:beta1} and \eqref{eq:beta2}.
    Different from Fig.~\ref{fig:BS_phig_dgy_duz_hole_total} in the main text, we set the center of mass momenta $\bm q =0$.
    The inset illustrates the presence of the BFS.
    }
    \label{fig:BS_phig_dgy_duz_hole_ap_beta2_total}
\end{figure}

\begin{figure}[htbp]
    \centering
    \includegraphics[width=0.5\textwidth]{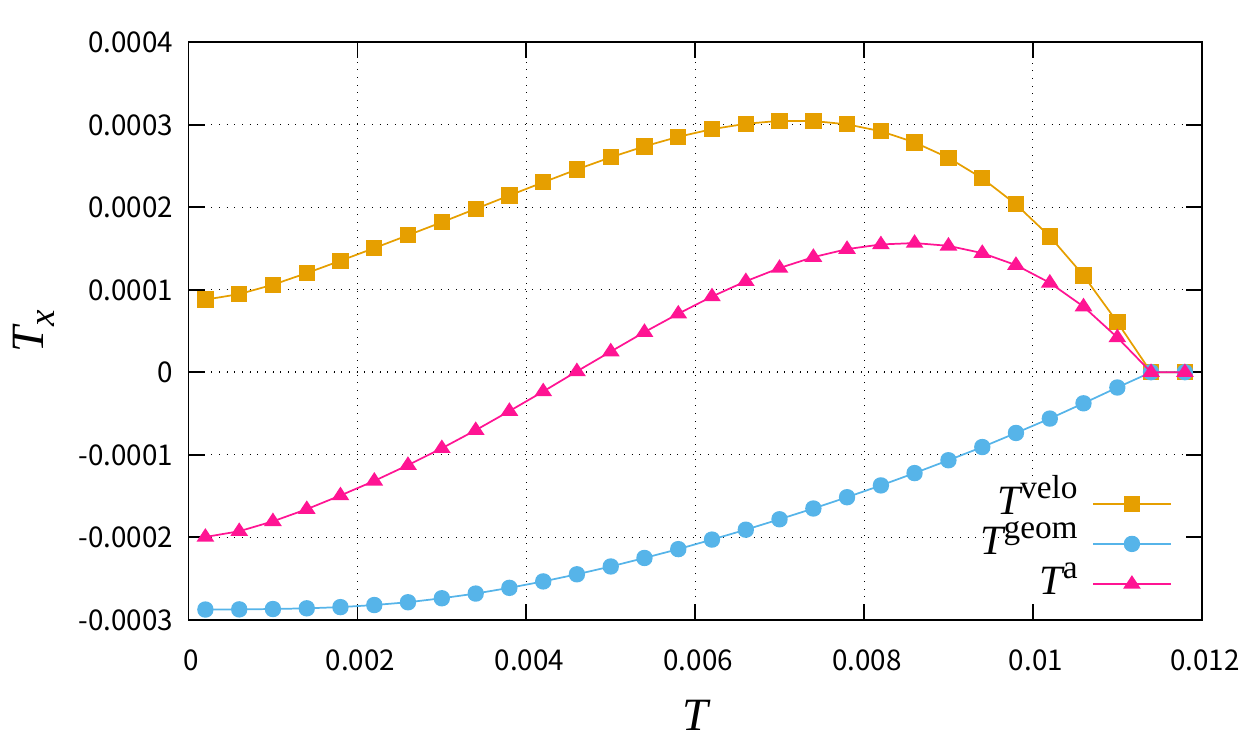}
    \caption{The temperature dependence of the anapole moment for the pair potential Eqs.~\eqref{eq:beta1} and \eqref{eq:beta2}.
    All colors show the same quantities as in Fig.~\ref{fig:ana_phig_duz_hole_ap}(a).}
    \label{fig:ana_phig_duz_hole_ap_beta05}
\end{figure}

On the other hand, when the system has the BFS, $f(E_{a\bm k})-f(E_{a-\bm k}) \neq 0$ even at $T=0$. Thus, we expect that the group velocity term does not completely disappear at $T=0$.
To verify this expectation, we consider the superconducting pair potential,
\begin{align}
    &\bm \Delta^{\rm g}(T) = \frac{1}{2}\Delta_{0}(T)\phi_{\rm g}^{0}\sigma_0\otimes\tau_0+\Delta_0(T)d_{{\rm g},y}^{z}\sigma_y\otimes\tau_z,\label{eq:beta1} \\
    &\bm \Delta^{\rm u}(T) = i\Delta_{0}(T)d_{{\rm u},z}^{0}\sigma_z\otimes\tau_0.\label{eq:beta2}
\end{align}
The difference from Eqs.~\eqref{eq:phig_dgy} and \eqref{eq:duz2} is only the factor $1/2$ in the first term of Eq.~\eqref{eq:beta1}.
We show the BS for $\bm q = 0$ in Fig.~\ref{fig:BS_phig_dgy_duz_hole_ap_beta2_total}.
Different from the cases discussed in Sec.~\ref{sec:UTe2}, the BFS appear even for $\bm q = 0$.
The temperature dependence of the anapole moment is shown in Fig.~\ref{fig:ana_phig_duz_hole_ap_beta05}, and indeed, we see that the group velocity term is not completely suppressed at $T=0$.
Thus, the presence of the BFS at $\bm q = 0$ enhances the group velocity term, consistent with the above expectation. In this case, the anapole moment at $T=0$ is not determined only by the geometric term. Even in this case, the sign change of the anapole moment can occur.

\section{Superfluid weight and center of mass momenta of Cooper pairs\label{appendix:sfw_qc}}

%We show the formula for the superfluid density.
Here, we consider the variation of free energy with respect to the center of mass momentum of Cooper pairs $q_{\mu}$ in a direction along which the anapole moment $T_{\mu}$ is finite.
Up to the second order of $q_\mu$, the superconducting free energy is expressed as,
\begin{eqnarray}
    F_{\bm q} = \dfrac{1}{2}D_{\mu\mu}^{\rm s}q^2_\mu + T_{\mu}q_\mu + F_{\bm 0}.
\end{eqnarray}
The superfluid density $D^{\rm s}_{\mu\mu}$ is given by the formula~\cite{kitamura2021superconductivity},
\begin{eqnarray}
    D_{\mu\mu}^{\rm s} &=& \dfrac{1}{2}\sum_{\bm k}\sum_{a}f(E_{a\bm k})\bra{\psi_{a\bm k}}\partial_{\mu}\partial_{\mu}H^-_{\bm k}\ket{\psi_{a\bm k}}\notag\\
    &+&\dfrac{1}{2}\sum_{\bm k}\sum_{ab}\dfrac{f(E_{a\bm k})-f(E_{a\bm k})}{E_{a\bm k}-E_{b\bm k}}\notag\\
    &\times&\bra{\psi_{a\bm k}}\partial_{\mu}H^+_{\bm k}\ket{\psi_{b\bm k}}\bra{\psi_{b\bm k}}\partial_{\mu}H^+_{\bm k}\ket{\psi_{a\bm k}},
\end{eqnarray}
where,
\begin{eqnarray}
	H^-_{\bm k} &=& \left(
		\begin{array}{cc}
			H_{\bm k}&0\\
			0&-H_{\bm k}
		\end{array}
	\right).
\end{eqnarray}
The superconducting free energy is rewritten as,
\begin{eqnarray}
    F_{\bm q} = \dfrac{1}{2}D_{\mu\mu}^{\rm s}\left(q_\mu + T_{\mu}/D_{\mu\mu}^{\rm s}\right)^2-T_{\mu}^2/2D_{\mu\mu}^{\rm s} + F_{\bm 0}.
\end{eqnarray}
Thus, the center of mass momentum $q_{\rm c}$ realizing the minimum free energy is estimated as $-T_{\mu}/D_{\mu\mu}^{\rm s}$. This formula is valid when $q_{\rm c}$ is small.
%We can attribute the sudden jump of $q_c$ to the differences of the $\Delta$ dependence between $T_{\mu}^{\rm a}$ and $D_{\mu\mu}^{\rm s}$.

%From the differences of the $\Delta$-dependence between $T_{\mu}^{\rm a}$ and $D_{\mu\mu}^{\rm s}$, we can understand the sudden jump of $q_c$.

\bibliography{main}

\begin{thebibliography}{80}
\bibitem [{jos()}]{josephcomment}%
  \BibitemOpen
  \href@noop {} {}\bibinfo {note} {The Josephson effect is proposed for the
  detection of the phase modulation in the helical
  superconductivity~\cite{kaur2005helical}. We consider that this method is
  also useful for observing the phase modulation in the anapole
  superconductivity.}\BibitemShut {Stop}%
\bibitem [{pau()}]{paulicomment}%
  \BibitemOpen
  \href@noop {} {}\bibinfo {note} {The two-orbital model with orbitals of same
  (opposite) parity is given by the permutation, $(\tau_x, \tau_y, \tau_z)
  \rightarrow (\tau_{z(y)}, \tau_{x(z)}, \tau_{y(x)})$}\BibitemShut {NoStop}%
\bibitem [{gro()}]{groupcomment}%
  \BibitemOpen
  \href@noop {} {}\bibinfo {note} {Considering that the group velocity term is
  also the first-order of $\partial \xi$, the sign of the $\partial \xi$, which
  is determined by structure of the band dispersion, is also crucial for the
  group velocity term.}\BibitemShut {Stop}%
\end{thebibliography}

\end{document}